\documentclass[prb,twocolumn,showpacs,preprintnumbers,amsmath,amssymb]{revtex4}
\usepackage{graphicx}
\usepackage{dcolumn}
\usepackage{bm}
\usepackage[dvips]{color}
\usepackage{bbm}           

\usepackage{float}
\usepackage{flafter}
\usepackage{epsfig}%

\newcommand{\blue}[1]{#1}
\begin{document}

\preprint{To Be Submitted}
\title{Displacement field and elastic constants in non-ideal crystals}

\author{C. Walz and M. Fuchs}
\affiliation{
Fachbereich f\"ur Physik, Universit\"at Konstanz, 78457 Konstanz, Germany}

\date{\today}

\begin{abstract}
In this work a periodic crystal with point defects is described in the framework of linear response theory for broken symmetry states using correlation functions and Zwanzig-Mori equations. The main results are microscopic expressions for the elastic constants and for the coarse-grained density, point-defect density, and displacement field, which are valid in real crystals, where  vacancies and interstitials are present. The coarse-grained density field differs from the small wave vector limit of the microscopic density. In the long wavelength limit, we recover the phenomenological description of elasticity theory including the defect density. 
\end{abstract}
\maketitle

\section{Introduction}

The theory of elasticity of solids started with Hooke in 1678, when he formulated the linear relation between stress and strain \cite{Hooke78}. The atomistic picture of matter contributed a quantitative microscopic understanding of the mechanical properties  of ideal crystals based on the particle potentials \cite{Born54}. Yet, the restriction to ideal crystals containing no point defects needs to be stressed. Nonequilibrium thermodynamics achieved a phenomenological description of the long wavelength and low frequency excitations \cite{Chaikin95,Forster75}. Martin, Parodi, and Pershan showed that the spontaneous breaking of continuous translational symmetry leads to eight hydrodynamic modes, one of which corresponds to defect diffusion \cite{Martin72}. Point defects, like vacancies and interstitials, are present in any equilibrium crystal and a complete microscopic theory of crystal dynamics needs to include them. Interestingly, such a complete microscopic theory of real crystals was lacking and is developed in this contribution in the framework of linear response and correlation functions theory of broken symmetry phases.

\begin{figure}[ht]
\begin{center}
\includegraphics[width=0.5\columnwidth]{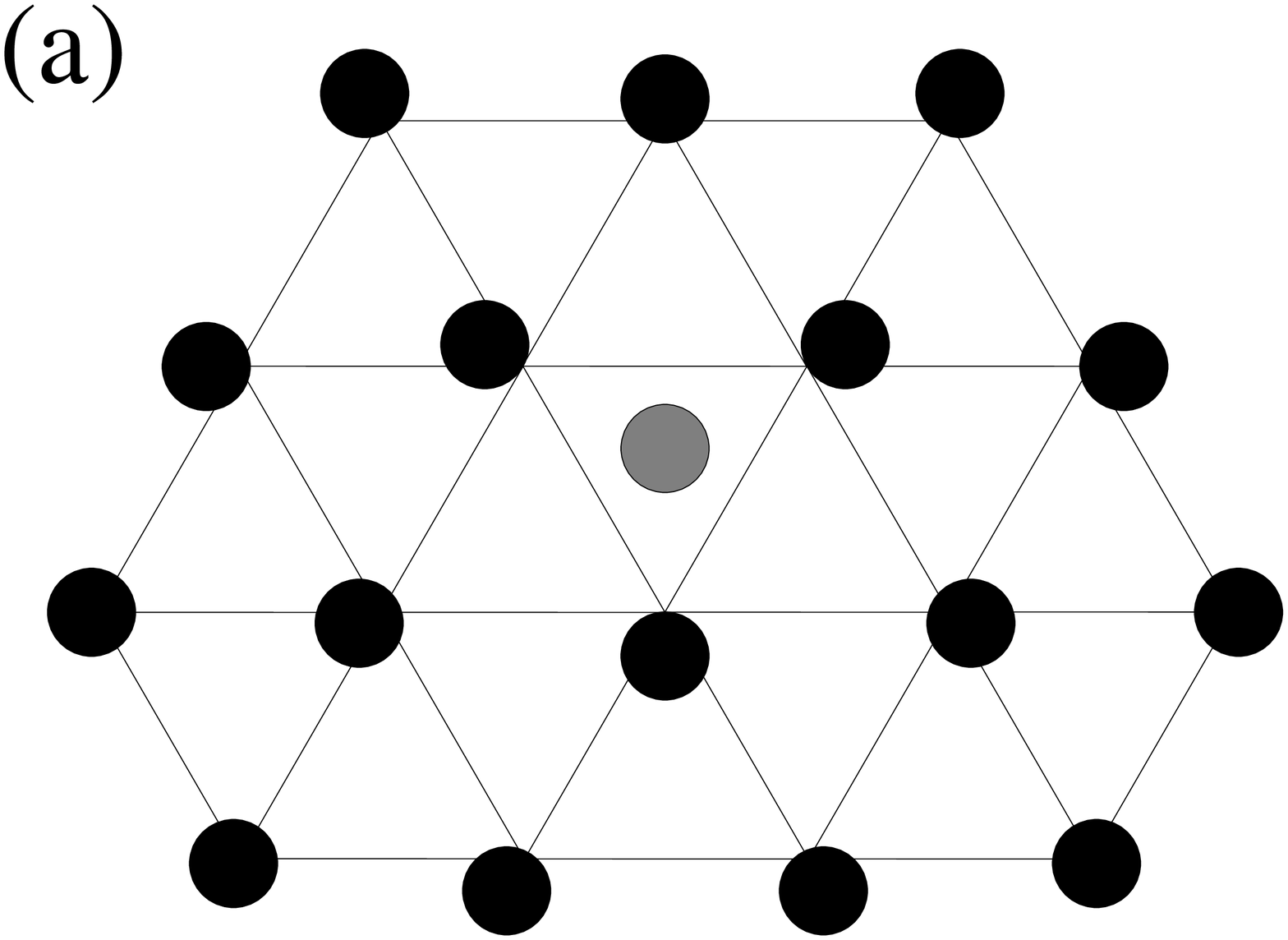}
\includegraphics[width=0.5\columnwidth]{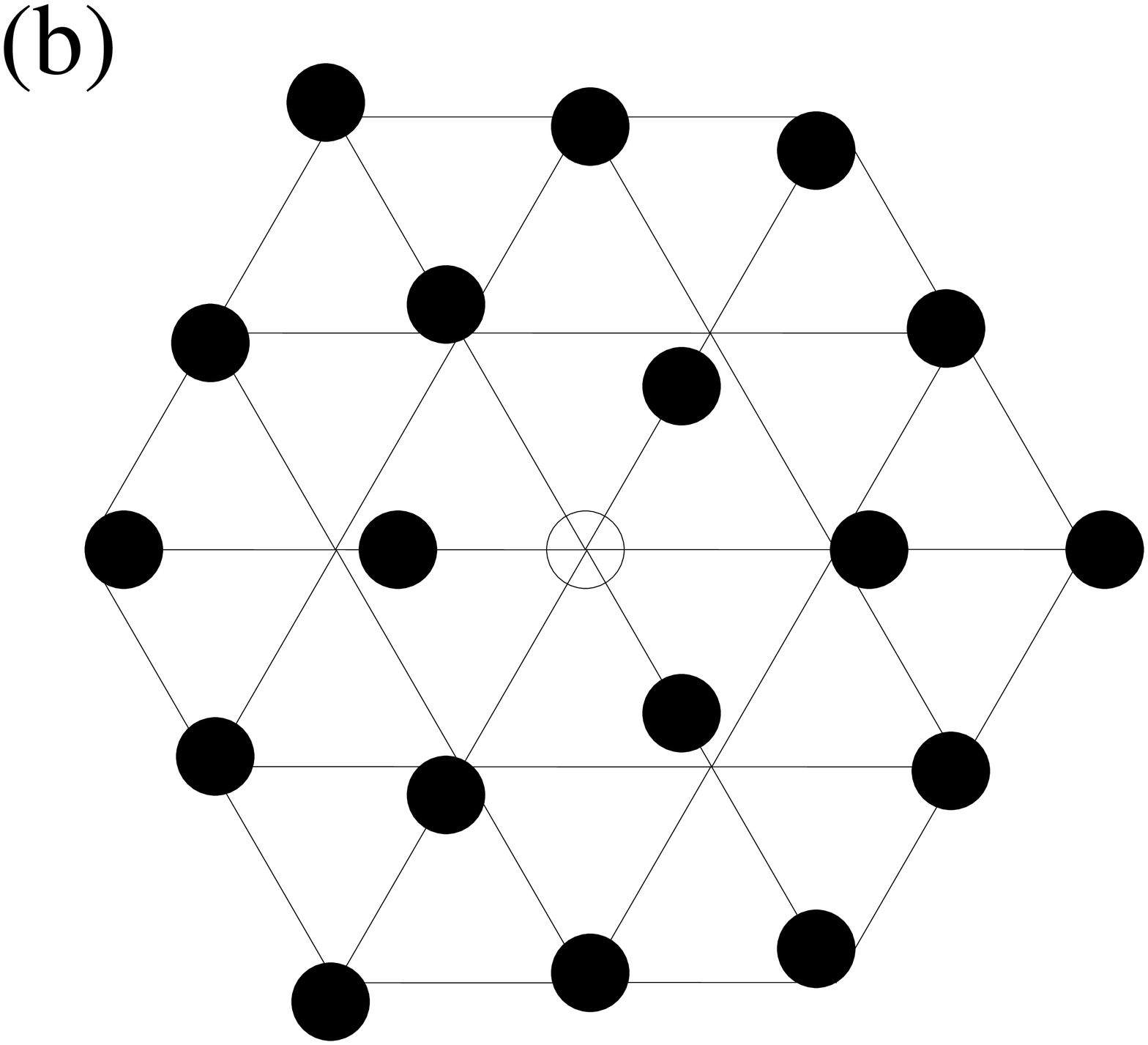}
\caption{\label{FigIntVac}Schematic two dimensional drawing of a crystal containing (a) an interstitial and (b) a vacancy with threefold symmetry\cite{Lechner08}.}
\end{center}
\end{figure}

Crystals exhibit long-range translational order and possess low-frequency Goldstone modes, e.g.~transverse sound waves, which try to restore the broken symmetry. In the familiar microscopic description of ideal crystals, the long-range order is incorporated at the start by assuming that the equilibrium positions of the particles are arranged in a perfect lattice. A one-to-one mapping follows between the particle $i$ and its lattice position $\mathbf{R}^i=\langle \mathbf{r}^i(t)\rangle$. The deviation between the actual position $\mathbf{r}^i(t)$ and the lattice position $\mathbf{R}^i$ is called displacement vector $\mathbf{u}^i(t)$
\begin{equation}
\mathbf{u}^i(t) = \mathbf{r}^i(t) -\mathbf{R}^i\qquad \mbox{(ideal crystal).}\nonumber
\end{equation}
The (symmetrized) gradient tensor of the displacement vector field is connected to the strain tensor, which plays the central role in the theory of elasticity. Yet, the applicability of the displacement vector is, due to the
one-to-one mapping, restricted to perfect crystals, because an interstitial corresponds to a particle without lattice site (Fig.~\ref{FigIntVac}a), and an vacancy to a lattice site without particle (Fig.~\ref{FigIntVac}b).
 Moreover, because defects are mobile, any 'improved' mapping would yield displacement vectors that can become arbitrarily large with time. Linear elasticity, considering small strain fields, thus would intrinsically be restricted to short times, contradicting/invalidating its application to low frequency vibrations. Thus the need arises to define the displacement field microscopically without the recourse to a perfect lattice of equilibrium sites $\mathbf{R}^i$.

In an ideal crystal where the one-to-one mapping of particles to lattice positions holds, a density change is given by the divergence of the displacement field  \cite{Fleming76,Landau89}
\begin{equation}
\delta n(\mathbf{r},t) = - n_0 \nabla \cdot \mathbf{u}(\mathbf{r},t)\qquad \mbox{(ideal crystal),}\nonumber
\end{equation}
with average density $n_0 =N/V$ and obvious definition of $\mathbf{u}(\mathbf{r},t)=\sum_i \mathbf{u}^i(t) \delta (\mathbf{r}-\mathbf{R}^i)$; note that it is a periodic lattice function in this case. Above relation holds because the density can only fluctuate by particles moving around their lattice sites.  In an ideal crystal, density  thus is not an independent degree of freedom, and description of the displacement field suffices. In a real crystal containing point defects, translational symmetry is still broken and long range order  exists, but the motion of defects decouples density fluctuations and the divergence of the displacement field. Density not only changes because of a deformation of the lattice, given by $-\nabla \cdot \mathbf{u}(\mathbf{r},t)$, but also by motion of additional/missing particles from one lattice cell to another. The density can be decomposed in two parts: 
\begin{equation}\label{relationdnuc}
\delta n(\mathbf{r},t) = - n_0 \nabla \cdot \mathbf{u}(\mathbf{r},t) - \delta c(\mathbf{r},t)\; .
\end{equation}
This definition of the defect density $c(\mathbf{r},t)$ is positive for vacancies, and negative for interstitials, i.e. the conventional sign favors the interpretation as a vacancy density.  Comparison with the discussion in [\onlinecite{Chaikin95}] for vacancies as the only point defects shows, that the thus defined variation in defect density is given in terms of number of vacancies $N_v$ and number of lattice sites $N_{ls}$
\begin{equation}
\delta c \leftrightarrow - \frac{N_{ls}}{V} \ \delta\!\left(\frac{N_v}{N_{ls}}\right)\; . \nonumber
\end{equation}
Thus, the magnitude of the variation of vacancy density increases, if there are more vacancies, and decreases, if there are more lattice sites.

Based on the relation \eqref{relationdnuc} alone, the hydrodynamic predictions by Martin {\it et al.}\cite{Martin72} can be recovered. Yet, there exists no microscopic particle based theory, which provides the definitions of the displacement and defect density fields, and recovers Eq.~\eqref{relationdnuc} from first principles. We will present these  definitions and derive the equations of motion for the fields, which reduce to the continuum description in the hydrodynamic limit. We will follow, within linear response theory, the accepted route  to symmetry broken states by considering conserved and symmetry restoring fields --- based on an application of Bogoliubov's inequality ---, followed by Zwanzig-Mori equations as pioneered by Kadanoff and Martin\cite{Kadanoff63}, and Forster\cite{Forster75}, and by taking the hydrodynamic limit at the end. 

In an important  contribution, Szamel and Ernst \cite{Szamel93} suggested the definition of the displacement field that we will find, which only uses density measurements without recourse to an underlying lattice. Importantly, the new expression for $\mathbf{u}(\mathbf{r},t)$ can thus be used for both ideal and real crystals, either by simulation or, experimentally, by optical techniques in e.g.~colloidal crystals.  Because we have a systematic way to discover the hydrodynamic fields and their equations, we can correct the  work by Szamel and Ernst and achieve consistency with the phenomenological description, which these authors could not \cite{Szamel93,Szamel97}. Our approach uses density functional theory (DFT), to describe the equilibrium correlations in a crystal, and thus superficially bears similarities to earlier works using approximate DFTs \cite{Ramakrishnan79,Jaric88,Jones87,Ryzhov95,Ferconi91,Mahato91,Velasco87,Xu88,Mazenko03}. In contrast to these previous works, we use exact DFT relations to simplify our expressions, and do not approximate the free energy functional, nor  start from parametrizations of density fluctuations \cite{Kirkpatrick90};
see Kirkpatrick {\it et al.}~\cite{Kirkpatrick90} for a discussion of these approximate theories, and  computer simulations \cite{Frenkel87,Jaric87,Velasco87,Runge87,Xu88} for possible problems arising concerning the elastic constants  \cite{Walz09}. 

The paper is organized as follows:
Section II derives the Zwanzig-Mori equations for (classical) crystalline solids, where translational symmetry is spontaneously broken and long range order exists. For simplicity, the set of conserved variables is restricted to density and momentum, neglecting energy. This restricts us to an isothermal approximation. Again for simplicity, memory kernels are neglected, restricting us to a dissipationless theory. Because the complete (infinite-dimensional) set of symmetry restoring variables, derived from Bogoliubov's inequality, is considered, a systematic approach to the dynamics of crystals is achieved; because we use the fluctuation dissipation theorem, a theory linearized close to equilibrium is obtained. Section III identifies the conventional fields used for describing the dynamics of crystals, especially the displacement and defect density field. Their equations of motion within the first Brillouin zone are derived. Section IV uses symmetry considerations within density functional theory, to derive the properties of the coefficients entering the equations of motion, and Sect. V discusses the results. First, the phenomenological equations of elasticity theory are recovered, and the elastic constants identified; we obtain their microscopic expressions in terms of the direct correlation functions of the crystal. Then the displacement and defect density field are discussed. Section VI ends the main text with short conclusions, and Appendix A shows consistency of the conventional but simplified Zwanzig-Mori equations of a crystal to our results.   

\section{General Theory}
\subsection{Microscopic Model and Microscopically Defined Hydrodynamic Variables}\label{microSect}

We consider a volume $V$ containing $N$ identical spherical particles at number density $n_0=N/V$. The motion of the particles with identical mass $m$ is described by a (classical) Liouville operator $\cal L$, which includes kinetic and (internal) potential energies. 

For the derivation of \blue{hydrodynamic} equations, the conserved quantities need to be considered. Starting with particle number, the (fluctuating) microscopic density is a sum over all particles $i$
\begin{equation}
\rho(\mathbf{r},t) = \sum_{i=1}^N \delta (\mathbf{r}-\mathbf{r}^i(t))\, .
\end{equation}
Temperature $T$ and density $n_0$ are chosen such, that the crystalline state gives the lowest free energy and translational invariance is spontaneously broken. Long-ranged order exists and the average density  varies periodically
\begin{equation}
n(\mathbf{r})= \langle \rho(\mathbf{r},t)\rangle =\sum_\mathbf{g} n_\mathbf{g} e^{i\mathbf{g\cdot r}}\; ,
\end{equation}
where the order parameters $n_\mathbf{g}$ are the Bragg-peak amplitudes
at the positions of the reciprocal lattice vectors $\bf g$, which are defined by
\begin{equation}
\mathbf{g\cdot L}= 2\pi n\; , 
\end{equation}
where $n$ is an integer, and $\mathbf{L}$ the set of discrete translational symmetry operations in real space. This means
\begin{equation}
n(\mathbf{r}) = n(\mathbf{r+L}) \quad \forall\ \mathbf{L}\; . \label{symm}
\end{equation}
An ensemble of identical crystals, which are just displaced in their
center of mass or overall orientation, yield vanishing order
parameters. To specify the broken symmetry state, it is thus
necessary to fix the six degrees of freedom of a rigid body
\cite{Mermin68}. Conceptually one describes the system in a frame of
the center of mass and orientation, or confines the crystal with the
help of external potentials. An example of such a potential is an
external wall, which in thermal equilibrium would need to be placed
such that $N_{ls}$ crystal lattice sites fit into the volume without
externally applied macroscopic strain or stress; here $N_{ls}$ differs
from $N$ because of point defects like vacancies and interstitials. 
The (canonical) ensemble, used to define the averages
$\langle\ldots\rangle=\int \hat{\rho}\ldots d\Gamma$, and the
corresponding $\cal L$ is henceforth restricted to contain such a
device which  fixes the degrees of freedom of a rigid body. Because
the internal fluctuations are not affected in the thermodynamic
limit, our results will depend on the canonical set of thermodynamic
variables (temperature $T$, number density $n_0$, volume $V$), and
the order parameters $n_{\bf g}$. Because they take their
equilibrium (non-strained) values, our (later) use of the
fluctuation-dissipation theorem restricts us to obtain the linear
equations of \blue{elasticity}, linearized around the equilibrium at
vanishing displacement field, $\langle{\bf u}\rangle\equiv0$.

The standard Fourier transformation in space  is used, where $d$ is the spatial dimension, and it will be stated explicitly if a specific spatial dimension is considered, which usually will be three dimensional space.
\begin{subequations}\label{microrhos}
\begin{align}
\rho(\mathbf{k},t) &= \int d^d\!r e^{-i\mathbf{k \cdot r}} \rho(\mathbf{r},t)=\sum_i^N e^{-i\mathbf{k\cdot r}^i(t)}\; ,\\
\rho(\mathbf{r},t) &= \int \frac{d^d\!k}{(2\pi)^d} e^{i\mathbf{k\cdot r}} \rho(\mathbf{k},t) \;.
\end{align}
\end{subequations}
Here the reciprocal vector $\mathbf{k}$ is unrestricted. The lattice symmetry also leads to periodicity in reciprocal space, which can be considered to be composed of periodically arranged Brillouin zones. If one restricts the reciprocal vector $\mathbf{q}$ to the first Brillouin zone, then the Fourier transformation of the density can be unambiguously decomposed into a reciprocal lattice vector $\mathbf{g}$ and $\mathbf{q}$
\begin{equation}
\rho(\mathbf{g+q},t) = \int d^d\!r e^{-i\mathbf{(g+q)\cdot r}} \rho(\mathbf{r},t)\; .
\end{equation}
The Fourier-back transformation simply becomes:
\begin{equation}
\rho(\mathbf{r},t) = \sum_{\bf g}  \int_{\rm 1^{st} BZ} \frac{d^d\!q}{(2\pi)^d}\; e^{i\mathbf{(g+q)\cdot r}}\; \rho(\mathbf{g+q},t) \;.
\end{equation}
This splitting of the Fourier coefficients of the density is useful as for the \blue{hydrodynamic description} one is interested in the long-wavelength fluctuations, i.e. $q\to 0$, close to all positions $\bf g$ of  the order parameters $n_\mathbf{g}$. Using the Fourier-transformed density the $n_\mathbf{g}$ are identified as
\begin{equation}\label{2dfnng}
n_\mathbf{g}=\frac{1}{V}\langle \rho(\mathbf{g})\rangle = \frac{1}{V}\sum_i^N \langle e^{-i\mathbf{g\cdot r}^i}\rangle\;.
\end{equation}

The second conserved quantity, to be considered is momentum.
For the momentum density $j_\alpha(\mathbf{r},t)$, which straightforwardly is given by
\begin{subequations}\label{micromomentums}
\begin{align}
j_\alpha(\mathbf{r},t) &= \sum_i^N p_\alpha^i \delta (\mathbf{r-r}^i(t))\;,\\
j_\alpha(\mathbf{k},t) &= \int d^d\!r e^{-i\mathbf{k\cdot r}}j_\alpha(\mathbf{r},t) = \sum_i^N p_\alpha^i e^{-i\mathbf{k\cdot r}^i(t)}\; ,
\end{align}
\end{subequations}
the distinction between $\mathbf{k}$ and $\mathbf{q}$ is not necessary. (Greek indices are used for spatial components, whereas latin ones denote particles.)

The conservation of particle density is expressed via (use of Einstein's sum convention is implied) 
\begin{equation}
 m \partial_{t} \rho (\mathbf{k},t) + ik_\alpha j_\alpha(\mathbf{k},t) = 0\;, 
\end{equation}
which follows from the microscopic definitions Eqs.~\eqref{microrhos} and \eqref{micromomentums}. The conservation of momentum density is stated through the divergence of the stress tensor.
\begin{equation}\label{consmomentum}
\partial_t j_\alpha(\mathbf{k},t) - i k_\beta \sigma_{\alpha\beta}(\mathbf{k}) =0 \;.
\end{equation}
A microscopic definition of the stress tensor can be found, for example, in [\onlinecite{Forster75}].
As third conserved field, the energy density should be considered. For simplicity however, we neglect the coupling of energy fluctuations to the mechanical fluctuations, restricting our results to an isothermal approximation. Extensions, incorporating energy fluctuations are straightforward, in principle.

\subsection{Bogoliubov Argument}

In a state with spontaneously broken symmetry, additional variables besides the conserved quantities need to be considered for deriving the \blue{continuum mechanics equations}. This by now classical route to \blue{'generalized hydrodynamic or elasticity theory'} --- in contrast to \blue{'hydrodynamic theory without broken symmetry'} --- builds on the Bogoliubov inequality to identify variables with long-ranged equilibrium correlations. For crystals this variant of Schwarz's inequality has been formulated by Wagner\cite{Wagner66}, 
\begin{equation}
\left\langle |\delta\rho(\mathbf{g}+\mathbf{q})|^{2} \right\rangle \ge \frac{\left|\left\langle j^\ast_\alpha({\mathbf{k}})\mathcal{L} \delta\rho (\mathbf{g}+\mathbf{q}) \right\rangle \right|^{2}}{\left\langle |\mathcal{L} j_\alpha({\mathbf{k}})|^{2}  \right\rangle}\;,
\end{equation}
where use is made of the hermitian property of $\langle (\mathcal{L}A)^\ast B\rangle$, and $\delta\rho$ denotes the density fluctuations from the equilibrium density
\begin{equation}\label{DfnDeltaRho}
\delta\rho(\mathbf{r},t) = \rho(\mathbf{r},t) - n(\mathbf{r})\;.
\end{equation}
The correlation functions required for the Bogoliubov inequality are considered in the following. Most of them will also be useful for elements of the so called frequency matrix.
First the classical equipartition theorem states for the correlation of the different spatial components of the momenta of the particles
\begin{equation}
\langle p^i_\alpha p^j_\beta \rangle = m k_{B}T \delta_{ij}\delta_{\alpha\beta}\;.
\end{equation}
With this
\begin{align}
\langle \delta j^\ast_\alpha(\mathbf{k},t) \delta j_\beta(\mathbf{k},t) \rangle = m n_0 Vk_BT \delta_{\alpha\beta}\; . \label{momentumcorr}
\end{align}
The standard properties of the Liouville operator \cite{Forster75}, here $\mathcal{L}=-i\partial_{t}$, yield
\begin{equation}
\mathcal{L}\delta\rho(\mathbf{g+q}) = \mathcal{L}\sum_i^N e^{-i(\mathbf{g+q})\cdot\mathbf{r}^i} = -\frac{(g+q)_\alpha}{m} j_\alpha(\mathbf{g+q})\; .
\end{equation}
Using Eq.~\eqref{2dfnng} the numerator of the Bogoliubov inequality becomes 
\begin{align}\label{jLrhocorr}
\left\langle j^\ast_\alpha({\mathbf{k}})\mathcal{L}\delta\rho(\mathbf{g}+\mathbf{q}) \right\rangle
 &=\! -\frac{(g+q)_\beta}{m}\left\langle j_\alpha^\ast({\mathbf{k}}) j_\beta(\mathbf{g}+\mathbf{q}) \right\rangle \notag\\
 &=\!  -(g+q)_\alpha  k_{B}T \left\langle\! \sum_{i}^N e^{-i(\mathbf{g}+\mathbf{q}-{\mathbf{k}})\cdot\mathbf{r}^{i}}\! \right\rangle \notag \\
 &=\! -(g+q)_\alpha k_{B}T V\, n_{\mathbf{g}+\mathbf{q}-{\mathbf{k}}}\; .
\end{align}
Thus $\bf k$ has to differ from $\bf q$ in the first Brillouin zone by a reciprocal lattice vector $\bf g'$ in order to give a finite order parameter $n_{\bf g-g'}$.
The denominator is expressed with the conservation of momentum density, Eq.~\eqref{consmomentum},
\begin{equation}\label{momentumR}
\left\langle |\mathcal{L}j_\alpha({\mathbf{k}})|^{2}  \right\rangle = {k}_\beta \langle \sigma_{\alpha\beta}\sigma_{\alpha\gamma}\rangle {k}_\gamma\;,
\end{equation}
or, neglecting directional dependence, \eqref{momentumR}$=R k^2$ with $R$ the correlation of the stress tensors.

So finally
\begin{align}
\left\langle |\delta\rho(\mathbf{g}+\mathbf{q})|^{2} \right\rangle \ge   \frac{ (\mathbf{g}+\mathbf{q})^{2}(k_{B}T)^{2} |n_{\mathbf{g}+\mathbf{q}-{\mathbf{k}}}|^{2}V^{2}}{R {k}^{2}}\;.
\end{align}
As $n_{\mathbf{g+q-{k}}}\ne 0$ only if $\mathbf{g+q-{k}={G}}$, with $\bf G$ a reciprocal lattice vector, the Bogoliubov inequality becomes
\begin{equation}
\left\langle |\delta\rho(\mathbf{{G}}+\mathbf{{k}})|^{2} \right\rangle \ge \frac{(\mathbf{{G}}+\mathbf{{k}})^{2}(k_{B}T)^{2} |n_{\mathbf{{G}}}|^{2}V^{2}}{R {k}^{2}}\propto {k}^{-2} \;.\label{bogol}
\end{equation}
Note that $\delta\rho(\mathbf{G}+\mathbf{k})$ is well defined for fixed $\mathbf{G}$, and in the hydrodynamic limit $\mathbf{k}\to 0$ can be replaced with $\mathbf{q}$, which lies in the first Brillouin zone.

Importantly the Bogoliubov inequality is an argument for all $\mathbf{G}\ne 0$, but not for $\mathbf{G}=0$, as in this case the right hand side of \eqref{bogol} is proportional to $k^0$. This component of the density is just conserved, it does not reflect the broken symmetry. Otherwise, at all finite reciprocal lattice vectors, the (expected) Bragg peak, which arises from the coherent scattering and is infinitely sharp in the present treatment because of the long-ranged order, sits  on top of a diverging (diffuse) background. Bogoliubov's inequality only gives a bound for the divergence for $k\to0$. In Sect.~\ref{sectionconstantselasticity} we will apply relations from density functional theory to prove the vanishing with $k^2$ of re-summed elements of the inverse of the density correlation functions, which corresponds to the equality sign in Eq.~(\ref{bogol}).

Following the standard reasoning to derive \blue{generalized elasticity theories}, the 'symmetry restoring' fluctuations need to be included in the set of slow variables and lead to Goldstone modes \cite{Forster75}. Equation (\ref{bogol}) shows that this requires to include the density fluctuations close to all reciprocal lattice vectors $\bf G$; note that we will apply Eq.~(\ref{bogol}) with the trivial notational change, $\bf G$ replaced by $\bf g$, in the following.

\subsection{Zwanzig-Mori Equations of Motion}

Whenever a set of slow variables $\{A_i(t)\}$ is selected, the Zwanzig-Mori formalism yields their linear equations of motion \cite{Forster75,Kubo95,Zwanzig01}. Neglecting dissipation, i.e. memory kernels, the 'reversible' equations of motion for small deviations $\langle \delta A_i(t)\rangle^{\rm lr}$ are given in terms of the equilibrium frequency matrix $\Omega_{ik}$
\begin{align}
\partial_t \langle \delta A_i(t)\rangle^{\rm lr} &= i \Omega^\ast_{ik} \langle \delta A_k(t)\rangle^{\rm lr} \label{1heom}\\
 &= i \Big(\langle \delta A_i^\ast \mathcal{L} \delta A_j\rangle \langle \delta A_j^\ast \delta A_k\rangle^{-1} \Big)^\ast \langle \delta A_k(t)\rangle^{\rm lr} \;. \notag
\end{align}
Here following Onsager and the fluctuation dissipation theorem, the deviations $\langle \delta A_i(t)\rangle^{\rm lr}$ of the specified variables from their equilibrium values are within linear response connected to correlation functions evaluated in the unperturbed system. The averages and the Liouville operator $\cal L$ defining the frequency matrix in Eq.~(\ref{1heom}) thus belong to the canonical ensemble introduced in Sect.~\ref{microSect}.

The equations of \blue{continuum mechanics} can be derived from
Eq.~(\ref{1heom}) by choosing as slow variables the
set of conserved and broken-symmetry restoring densities, and then
analyzing the limit of small wave vectors, $q\to0$ . Based on
Bogoliubov's inequality Eq.~(\ref{bogol}), the set of variables
comprises the $d$ components of the conserved momentum density $\delta
j_\alpha(\mathbf{q},t)$, and the Fourier components of the density
fluctuations $\delta \rho(\mathbf{g+q},t)$ close to the Bragg-peak
positions; to uniquely specify the latter, let us recall that the
wave vector $\bf q$ is restricted to lie in the first Brillouin zone.
To clarify the  notation in the following, we abbreviate
\begin{subequations}
\begin{align}
\delta n_\mathbf{g}(\mathbf{q},t) &= \langle \delta \rho(\mathbf{g+q},t)\rangle^{\rm lr}\;,\\
\delta j_\alpha (\mathbf{q},t) &= \langle \delta j_\alpha (\mathbf{q},t) \rangle^{\rm lr}\; .
\end{align}
\end{subequations}
\blue{Zwanzig-Mori's equations} (\ref{1heom}) will for this choice of variables in the limit of small wave vector lead to the dissipationless, isothermal (i.e. neglecting coupling to heat flow), and linearized equations of \blue{crystal elasticity}.

Most of the  elements of the frequency matrix $\Omega_{ik}$ have been derived in the previous section, namely in Eqs. (\ref{momentumcorr}) and (\ref{jLrhocorr}); note that the latter will be used in the following for wave vectors ${\bf k}={\bf q}$ in the first Brillouin zone only.
Many matrix elements vanish because of symmetry \cite{Berne00}. In the problem at hand the most useful symmetry in this respect is invariance under time reversal, as the dynamical variables have a definite parity (even for the Fourier components of the density and odd for the momentum density), as well as the Liouvillian $\mathcal{L}$ (odd).

The only non-vanishing matrix element still missing is the inverse $J_\mathbf{g g^{\prime}}$ of the density correlation matrix,
which is defined by
\begin{equation}\label{invrhocorr}
Vk_BT \delta_\mathbf{g g^{\prime\prime}} = \sum_\mathbf{g^\prime} \langle \delta\rho^\ast(\mathbf{g+q})\delta\rho(\mathbf{g^\prime+q}) \rangle J_\mathbf{g^\prime g^{\prime\prime}}(\mathbf{q})\; .
\end{equation}
Its properties will be discussed in Sect.~\ref{symmDensCorFnc}.

Thus the dissipationless and isothermal \blue{Zwanzig-Mori equations of motion} of a crystal are
\begin{subequations}\label{bsheom}
\begin{align}
\partial_t  \delta n_\mathbf{g}(\mathbf{q},t) &=  i \left(\frac{\langle \delta\rho (\mathbf{g+q})^\ast\mathcal{L}\delta j_\alpha(\mathbf{q})\rangle}{\langle \delta j_\alpha^\ast(\mathbf{q}) \delta j_\beta(\mathbf{q}) \rangle}\right)^{\!\!\ast} \delta j_\beta(\mathbf{q},t) \notag \\
 &= -i\frac{n_\mathbf{g}}{mn_0}(g+q)_\alpha \ \delta j_\alpha(\mathbf{q},t)\;, \label{ngheom}\\
\partial_t \delta j_\alpha(\mathbf{q},t) &=  i \sum_\mathbf{g^\prime ,g}\left( \frac{\langle \delta j_\alpha(\mathbf{q})^\ast \mathcal{L}\delta\rho(\mathbf{g^\prime+q})\rangle}{\langle\delta\rho^\ast(\mathbf{g^\prime +q}) \delta\rho(\mathbf{g+q})\rangle}\right)^{\!\!\ast}  \delta n_\mathbf{g}(\mathbf{q},t) \notag \\
 &= -i\sum_{\mathbf{g^\prime},\mathbf{g}} (g^\prime+q)_\alpha n^\ast_\mathbf{g^\prime} J^\ast_{\mathbf{g^\prime g}}(\mathbf{q}) \ \delta n_\mathbf{g}(\mathbf{q},t)\;. \label{jheom}
\end{align}
\end{subequations}
Although formally exact, these equations still need a lot of interpretation. To begin with, there are $\infty +3$ of them in three dimensions. Naturally the question arises how the set of $\{\delta n_\mathbf{g}(\mathbf{q},t), \delta j_\alpha(\mathbf{ q},t)\}$, in the limit of small wave vector $q$, reduces to the seven conventional ones of elasticity theory $\{\delta n(\mathbf{q},t), \delta u_\alpha(\mathbf{q},t), \delta j_\alpha(\mathbf{q},t)\}$, if the coarse-grained density $\delta n(\mathbf{q},t)$ (or instead the vacancy density $\delta c(\mathbf{q},t)$) and the displacement field $\delta u_\alpha(\mathbf{q},t)$ are used, as in the case of phenomenological theory\cite{Chaikin95,Martin72,Fleming76} (see Sec.~\ref{phenomenology} for a summary of phenomenology). In terms of the frequency matrix this corresponds to solving the eigenvalue problem, thus showing that this matrix has seven eigenvalues which become arbitrarily small in the limit $q\to0$. These eigenvalues are the ones of \blue{classical elasticity theory}, and their corresponding eigenvectors are the variables of the continuum approach derived within our microscopic theory.

\section{Relation to Classical Elasticity}\label{2identdisplacement}
The \blue{Zwanzig-Mori} equations of motion \eqref{bsheom} can be written with the frequency matrix in a compact notation, in order to analyze the hydrodynamic limit. Before doing this in general, the wave equation, which contains the constants of elasticity or sound velocities, can be read off immediately.

\subsection{Wave Equation}
Taking a time derivative of Eq.~(\ref{jheom}), and combining it with
Eq.~(\ref{ngheom}), leads to a closed equation of motion for the momentum density
\begin{align}
&\partial^2_t \delta j_\alpha(\mathbf{q},t) = &\notag \\ &-\frac{1}{mn_0}\!\sum_{\mathbf{g^\prime,g}}\!(g^\prime +q )_\alpha n_\mathbf{g^\prime}^\ast J_{\!\mathbf{g^\prime\!g}}^\ast(\mathbf{q}) n_\mathbf{g}(g+q)_\beta \delta j_\beta(\mathbf{q},t)& \notag\\
& = -\frac{1}{mn_0} \Lambda_{\alpha\beta}(\mathbf{q}) \ \delta j_\beta(\mathbf{q},t)\; .&\label{2wavemomentum}
\end{align}
This equation can take the required form of the wave equation, if the  $d\times d$-dimensional matrix $\Lambda_{\alpha\beta}(\mathbf{q})$ vanishes quadratically with wave vector going to zero, $\Lambda_{\alpha\beta}(\mathbf{q}) ={\cal O}(q^2)$ for $q\to0$. This property and the relation with the elastic constants, which obey the Voigt symmetry in their indices, is the subject of chapter \ref{chapterelasticity}. Strictly speaking, only then  the term wave equation is justified. From Eq.~(\ref{2wavemomentum}) one reads off
\begin{align}
\Lambda_{\alpha\beta}(\mathbf{q}) &= \sum_{\mathbf{g^\prime,g}} (g^\prime +q )_\alpha n_\mathbf{g^\prime}^\ast J_{\mathbf{g^\prime g}}^\ast(\mathbf{q}) n_\mathbf{g}(g+q)_\beta\; . \label{Lambdadfn}
\end{align}
The remarkable feature of this equation, however, is that it is exact and holds for wave vector $\bf q$ throughout the  first Brillouin zone. It is independent of the yet to be found relation between $\delta n_\mathbf{g}(\mathbf{q},t)$ with the displacement field $\delta u_\alpha(\mathbf{q},t)$ and the defect density $\delta c(\mathbf{q},t)$. The only input for this relation are the exact matrix elements of $\langle \delta j^\ast_\alpha(\mathbf{q})\delta j_\beta(\mathbf{q})\rangle$, $\langle j^\ast_\alpha(\mathbf{q})\mathcal{L}\delta\rho(\mathbf{g+q})\rangle$, and $\langle \delta\rho^\ast(\mathbf{g+q}) \delta\rho(\mathbf{g^\prime+q})\rangle$.\\

\subsection{Displacement Field and Defect Density}\label{subsecDFDD}

The \blue{Zwanzig-Mori equations of motion} \eqref{bsheom} of the set of conserved and Goldstone modes couple density fluctuations with modulation given by (almost) the reciprocal lattice vectors $\bf g$. To bring out the contributions from the various $\bf g$, consider $\delta \vec{n}(\mathbf{q},t)$ as infinite-dimensional (column) vector, whose components $\delta n_{\bf g}(\mathbf{q},t)$ are indexed by the $\bf g$ (ordered in some fixed but arbitrary fashion). Let  $v_{\alpha,{\bf g}}=i(g+q)_\alpha n_\mathbf{g}$ be an element of a constant infinite-dimensional vector $\vec{v}_\alpha$ (index $\alpha$ continues to label the spatial coordinate), and let $\stackrel{\leftrightarrow}{J}= J_{\mathbf{g g^\prime}}$ be a corresponding $\infty\times\infty$-dimensional  matrix.
Thus the \blue{Zwanzig-Mori equations} \eqref{bsheom} take the form
\begin{subequations}\begin{align}
\partial_t \delta\vec{n}(\mathbf{q},t) &= -\frac{1}{mn_0} \vec{v}_\alpha\; \delta j_\alpha(\mathbf{q},t)\;,\\
\partial_t \delta j_\alpha(\mathbf{q},t) &= \vec{v}^{\,\dagger}_\alpha \stackrel{\leftrightarrow}{J}^\ast\delta \vec{n}(\mathbf{q},t)\;,
\end{align}\end{subequations}
with $\vec{v}^{\,\dagger}_\alpha = -i(g+q)_\alpha n_\mathbf{g}^\ast$ an infinite-dimensional (row) vector, and $\stackrel{\leftrightarrow}{J}^\ast\delta \vec{n}$ a shorthand notation for $\sum_{\bf g'} J_{\bf g g'}^\ast \delta n_{\bf g'}$, etc.

The prequel of the wave equation, Eq.~(\ref{2wavemomentum}), of the momentum density immediately is recovered and takes the form
\begin{align}
\partial_t^2 \delta j_\alpha(\mathbf{q},t) = \frac{-1}{mn_0}\vec{v}^{\,\dagger}_\alpha\! \stackrel{\leftrightarrow}{J}^\ast\!\!\vec{v}_\beta  \delta j_\beta(\mathbf{q},t)
 = \frac{-1}{mn_0} \Lambda_{\alpha\beta} \delta j_\beta(\mathbf{q},t) .
\end{align}

In a perfect crystal, density fluctuations result from  the divergence of the displacement field\cite{Fleming76,Landau89}
\begin{equation}
\delta n(\mathbf{q},t) = -in_0 q_\alpha \delta u_\alpha(\mathbf{q},t)\; ,
\end{equation}
motivating the following defining relation
\begin{equation}
\delta\vec{n}(\mathbf{q},t) = -\vec{v}_\alpha\delta u_\alpha(\mathbf{q},t)\; ,
\end{equation}
for the (Fourier transformed) displacement field $\delta u_\alpha(\mathbf{q},t)$. Equation (\ref{ngheom}) then becomes
\begin{equation}
\vec{v}_\alpha \left\{\partial_t\delta u_\alpha(\mathbf{q},t)-\frac{1}{mn_0}\delta j_\alpha(\mathbf{q},t) \right\} = 0\; ,
\end{equation}
which states for an ideal crystal, as expected,
that the time derivative of the displacement is the velocity field, i.e.~the momentum density field divided by the mass density
\begin{equation}\label{velocity}
\partial_t\delta u_\alpha(\mathbf{q},t) = \frac{1}{mn_0}\delta j_\alpha(\mathbf{q},t) \;.
\end{equation}

Generalizing this consideration in  a real crystal, the (Fourier transformed) defect density field $\delta c(\mathbf{q},t)$ can be defined by the difference between the density fluctuations and the divergence of the displacement field
\begin{subequations}
\begin{equation}
\delta \vec{n}(\mathbf{q},t) = -\vec{v}_\alpha \delta u_\alpha(\mathbf{q},t) + \vec{v}^{\,\prime} \delta c(\mathbf{q},t)\;, \label{deltac}
\end{equation}
with $\vec{v}^{\,\prime}$ some yet unknown constant vector with components $v'_{\bf g}$.
For convenience this can be rewritten with $\vec{v}_\alpha^{\,0}=\vec{v}_\alpha(\mathbf{q}=0)$, whose components $\vec{v}_\alpha^{\,0}=ig_\alpha n_\mathbf{g}$ are the Fourier components of the gradient of the equilibrium density.
\begin{equation}
\delta \vec{n}(\mathbf{q},t) = -\vec{v}_\alpha^{\,0}\delta u_\alpha(\mathbf{q},t) -i q_\alpha\vec{n}\delta u_\alpha(\mathbf{q},t) + \vec{v}^{\,\prime} \delta c(\mathbf{q},t)\; ,
\end{equation}
\end{subequations}
with $\vec{n}$ the vector with components $n_{\bf g}$. It is an ansatz that the four (i.e.~$d+1$) dynamical variables introduced, namely the $\delta u_\alpha(\mathbf{q},t)$ and $\delta c(\mathbf{q},t)$, together with the three ($d$) familiar components $\delta j_\alpha(\mathbf{q},t)$, solve the infinite set of equations \eqref{bsheom}, whose uniqueness we can not prove. Yet, the proof that this ansatz solves Eqs.~\eqref{bsheom}, is straightforward and leads to relations expected from phenomenology (see Sec.~\ref{phenomenology}). Entering  Eq.~\eqref{deltac} into Eq.~(\ref{ngheom}) and decomposing $\vec{v}_\alpha= \vec{v}_\alpha^{\,0} + i q_\alpha \vec{n}$, one arrives at
\begin{align}
&-\vec{v}_\alpha^{\,0}  \left\{\partial_t\delta u_\alpha (\mathbf{q},t) -\frac{1}{mn_0}\delta j_\alpha(\mathbf{q},t)\right\} \notag\\
&=\!\vec{n}\left\{\!-\frac{i}{mn_0} q_\alpha \delta j_\alpha(\mathbf{q},t) +i q_\alpha \partial_t\delta u_\alpha(\mathbf{q},t)\!\right\}\!-\vec{v}^{\,\prime}\partial_t\delta c(\mathbf{q},t) \; .\notag
\end{align}
If one continues to require that the time derivative of the displacement is the velocity field, Eq.~\eqref{velocity}, then the first bracket vanishes as before. As for the general case with dissipation $\partial_t \delta c(\mathbf{q},t)\ne 0$ one needs a further relation for the vanishing of the right hand side for all $q$. This is achieved by taking the unknown constant vector $\vec{v}^{\,\prime} = -\vec{n}/n_0$. One is then able to define the coarse-grained density variation by the expected relation \eqref{relationdnuc}
\begin{equation}\label{deltan}
\delta n(\mathbf{q},t)= -i n_0q_\alpha \delta u_\alpha(\mathbf{q},t)- \delta c(\mathbf{q},t)\; ,
\end{equation}
which states that density fluctuations are composed of the divergence of the displacement field and defect density fluctuations. Consequently,  the original Eq.~(\ref{ngheom}) is solved for all $q$ by the conservation law of mass or particle number, which causes also the second bracket to vanish
\begin{align}
&-\vec{v}_\alpha^{\,0} \left\{\underbrace{\partial_t\delta u_\alpha(\mathbf{q},t)-\frac{1}{mn_0}\delta j_\alpha(\mathbf{q},t)}_{=0}\right\}\notag\\
 &= -\vec{n}\left\{\underbrace{\frac{1}{n_0}\partial_t\delta n(\mathbf{q},t) +\frac{i}{mn_0}q_\alpha\delta j_\alpha(\mathbf{q},t)}_{=0} \right\}\; .\notag
\end{align}
Equation (\ref{ngheom}) thus is solved by the ansatz for all $\bf q$ in the first Brillouin zone.

Turning to the second \blue{Zwanzig-Mori equation} (\ref{jheom}), the ansatz Eq.~(\ref{deltac}) transforms it into
\begin{align}
&\partial_t \delta j_\alpha(\mathbf{q},t) = -\vec{v}^{\,\dagger}_\alpha\stackrel{\leftrightarrow}{J}^\ast\vec{v}_\beta \delta u_\beta(\mathbf{q},t)- \vec{v}^{\,\dagger}_\alpha\stackrel{\leftrightarrow}{J}^\ast\frac{\vec{n}}{n_0}\delta c(\mathbf{q},t)\notag\\
&= - \Lambda_{\alpha\beta}  \delta u_\beta(\mathbf{q},t) - V_\alpha \delta c(\mathbf{q},t)\; ,\label{zwischen}
\end{align}
with constant $d$-dimensional vector $V_\alpha$ given by
\begin{align}
& V_\alpha({\bf q}) = \vec{v}^{\,\dagger}_\alpha\stackrel{\leftrightarrow}{J}^{\ast}\!\!\frac{\vec{n}}{n_0} =-\frac{i}{n_0} \sum_\mathbf{g^{\prime},g} (g^{\prime}+q)_\alpha n_\mathbf{g^{\prime}}^\ast J_\mathbf{g^{\prime}g}^\ast(\mathbf{q}) n_\mathbf{g}\; .\label{couplingc}
\end{align}
This equation is consistent with the wave equation for the momentum density, when a time derivative is taken, Eq.~\eqref{velocity} is used, and the defect density is  constant in time
\begin{equation}
-V_\alpha \partial_t\, \delta c(\mathbf{q},t)= \partial_t^2 \delta j_\alpha(\mathbf{q},t) + \frac{1}{mn_0} \Lambda_{\alpha\beta} \delta j_\beta(\mathbf{q},t) =  0 \; .
\end{equation}
Alternatively, Eq.~(\ref{zwischen}) leads to the wave equation for the displacement field when the time derivative of the displacement is again identified as velocity field
\begin{equation}\label{uwaveeqn}
\partial_t^2 \delta u_\alpha(\mathbf{q},t)\!=\!-\frac{1}{mn_0} \Lambda_{\alpha\beta}(\mathbf{q}) \delta u_\beta(\mathbf{q},t)
  \!-\! \frac{1}{mn_0} V_\alpha({\bf q}) \delta c(\mathbf{q},t) .
\end{equation}
Importantly, the matrix of elastic coefficients $\Lambda_{\alpha\beta}({\bf q})$ in this prequel of the wave equation is identical to the one derived for the momentum density, Eq.~\eqref{2wavemomentum}. Equation \eqref{uwaveeqn} can thus reduce to the expected result from \blue{classical elasticity theory} in the limit of small wave vector, if $V_\alpha({\bf q})={\cal O}(q)$ can be shown for $q\to 0$. This will be discussed together with the properties of $\Lambda_{\alpha\beta}$ in Sect.~\ref{sectionconstantselasticity}. Then we will be able to conclude that the \blue{Zwanzig-Mori equations of a crystal} \eqref{bsheom} are solved by seven ($2d+1$) coarse-grained fields which are the momentum density $\delta j_\alpha(\mathbf{q},t)$, and the displacement $\delta u_\alpha(\mathbf{q},t)$ and defect density $\delta c(\mathbf{q},t)$ field introduced in Eq.~\eqref{deltac}. We will also be able to conclude that the coarse grained fields, whose equations of motion were just determined in this section for $\bf q$ in the first Brillouin zone, obey the equations of motion known from phenomenological elasticity theory in the limit of $q\to0$. In support of this, Eq.~\eqref{uwaveeqn} finds that the prefactors in front of $\delta u_\beta(\mathbf{q},t)$ and $\delta c(\mathbf{q},t)$ are connected. Their $q\to0$-limits  reduce to thermodynamic derivatives, which, from equilibrium thermodynamics, have to satisfy Maxwell relations \cite{Martin72,Fleming76}. While the relations are not sufficient to express one coefficient in turn of the other, the connection between $\Lambda_{\alpha\beta}$ and $V_\alpha$ in Eq.~\eqref{uwaveeqn} closely mirrors the thermodynamic one expected in \blue{classical elasticity theory}; see Sect.~\ref{phenomenology}.

The question remains how $\delta u_\alpha(\mathbf{q},t)$ and $\delta c(\mathbf{q},t)$, given implicitly in Eq.~\eqref{deltac}, can be obtained directly in terms of the density fluctuations $\delta n_{\bf g}({\bf q},t)$. Fortunately, this can be achieved by projecting $\delta \vec{n}({\bf q},t)$ onto the two vectors $\vec{v}_\alpha^{\,0}$ and $\vec{n}$, as they are orthogonal
\begin{equation}\label{SumsOrthogonal}
i \; \left(\vec{v}_\alpha^{\,0}\right)^\dagger \; \vec{n} = \sum_\mathbf{g} |n_\mathbf{g}|^2 g_\alpha =0\;,
\end{equation} 
because of symmetry.
Projecting the ansatz for the density fluctuations in Eq.~\eqref{deltac} onto $\vec{v}_\alpha^{\,0}$ gives an explicit formula for the displacement field $\delta u_\alpha(\mathbf{q},t)$ in terms of the $\delta n_\mathbf{g}(\mathbf{q},t)$
\begin{equation}\label{firstsummation}
\delta u_\alpha(\mathbf{q},t) = i\mathcal{N}^{-1}_{\alpha\beta}\sum_\mathbf{g} n_\mathbf{g}^\ast\, g_\beta \; \delta n_\mathbf{g}(\mathbf{q},t)\;,
\end{equation}
with $\mathcal{N}_{\alpha\beta} = \sum_\mathbf{g} |n_\mathbf{g}|^2 g_\alpha g_\beta$.
A second summation over $\delta n_\mathbf{g}(\mathbf{q},t)$ obtained from projecting Eq.~\eqref{deltac} on $\vec n$ yields the hydrodynamic variation of the coarse-grained density
\begin{align}\label{secondsummation}
\delta n(\mathbf{q},t) &= \frac{n_0}{\mathcal{N}_0}\sum_\mathbf{g} n_\mathbf{g}^\ast \delta n_\mathbf{g}(\mathbf{q},t)\;,
\end{align}
where $\mathcal{N}_0=\sum_\mathbf{g} |n_\mathbf{g}|^2$. With \eqref{deltan}, the relation  between the variation of coarse-grained density, the lattice density, and the defect density, we get
\begin{align}\label{2hydrodefectdensity}
\delta c(\mathbf{q},t) = -n_0 \sum_\mathbf{g} n_\mathbf{g}^\ast\Big( \frac{1}{\mathcal{N}_0} -  \mathcal{N}_{\alpha\beta}^{-1}q_\alpha g_\beta\Big) \delta n_\mathbf{g}(\mathbf{q},t) \; .
\end{align}

\section{Symmetry and Invariance}\label{chapterelasticity}
This chapter completes the derivation of \blue{Zwanzig-Mori equations of motion} in terms of the Fourier components of the density by showing that the proper characteristics are recovered in the hydrodynamic limit.
Properties of the (inverse) density correlation matrix lead,
due to translational invariance, to the correct $q$ dependence and, due to rotational invariance, to symmetries of the constants of elasticity.

\subsection{Symmetries of Density Correlation Functions}\label{symmDensCorFnc}

The symmetry property in Eq.\eqref{symm} of the average density of a crystal is very familiar.  Also important is the symmetry property of the equilibrium two-point correlation function \cite{McCarley97}. For example the correlation of the density fluctuations
$\delta \rho(\mathbf{r},t) = \rho(\mathbf{r},t) - n(\mathbf{r})$
is also periodic
\begin{equation}
\langle \delta \rho(\mathbf{r_1}) \delta\rho(\mathbf{r_2})\rangle = \langle \delta \rho(\mathbf{r_1+L}) \delta\rho(\mathbf{r_2+L})\rangle \quad \forall\ \mathbf{L}\;.
\end{equation}
This results in a periodic center of mass variable ${\bf R}=\frac{\mathbf{r_1+r_2}}{2}$ and a Fourier coefficient, which depends on the difference $\Delta{\bf r}=\mathbf{r_1-r_2}$
\begin{equation}\label{twopointcrystal}
\langle \delta \rho(\mathbf{r_1})\delta \rho(\mathbf{r_2})\rangle = \sum_\mathbf{g} e^{i\mathbf{g}\cdot\frac{\mathbf{r_1+r_2}}{2}} \delta n^{(2)}_\mathbf{g}(\mathbf{r_1-r_2})\;.
\end{equation}
As the density is a real quantity and the correlation function is symmetric with respect to interchange of variables, the $\delta n^{(2)}_\mathbf{g}(\mathbf{r_1-r_2})$ obey the following two equations
\begin{align}
\delta n^{(2)}_\mathbf{g}(\mathbf{r_1-r_2}) = \delta n^{(2)}_\mathbf{-g}(\mathbf{r_1-r_2})
 = \delta n^{(2)}_\mathbf{g}(\mathbf{r_2-r_1})\;.
\end{align}
Rewriting Eq.~\eqref{twopointcrystal}, one realizes that the Fourier
transformation of the Fourier coefficient $\delta
n^{(2)}_\mathbf{g}(\Delta\mathbf{r})$ with respect to the difference
coordinate $\Delta\mathbf{r}$ can be understood as a generalized
structure factor $S_\mathbf{g}(\mathbf{k})$
\begin{subequations} \label{2GSF}
\begin{align}
S_\mathbf{g}(\mathbf{k}) &= \frac 1V \int d^d\!r_1\!\int d^d\!r_2 \ \langle \delta\rho(\mathbf{r_1})\delta\rho(\mathbf{r_2})\rangle\ e^{-i\mathbf{g\cdot R}} \ e^{-i\mathbf{k\cdot \Delta r}}\\
&= \int d^d\!\Delta r \ \delta n^{(2)}_\mathbf{g}(\mathbf{\Delta r})e^{-i\mathbf{k\cdot \Delta r}}\\
&=  \delta n^{(2)}_\mathbf{g}(\mathbf{k}) = \frac 1V \langle \delta\rho(\mathbf{g/2 +k}) \delta\rho(\mathbf{g/2 -k}) \rangle\;. \label{symSg3}
\end{align}
\end{subequations}
The generalized structure factor $S_\mathbf{g}(\mathbf{k})$ is, due
to the symmetry of the crystal, a density fluctuation function
evaluated with a combination of a reciprocal lattice vector
$\mathbf{g}$ and a reciprocal vector $\mathbf{k}$. Bogoliubov's
inequality shows that $S_0(k)$ diverges quadratically at all
reciprocal lattice vectors $\mathbf{k\to \tilde{g}}\ne 0$, which however is not enough
information to simplify completely the expressions for the elastic
coefficients that we derived in the previous
Sect.~\ref{2identdisplacement}. It remains to study the complete
matrix of inverse density correlations, $J_\mathbf{g
g^\prime}(\mathbf{q})$ defined in Eq.~(\ref{invrhocorr}), which we
undertake now  using density functional theory \cite{Evans79,
Rowlinson82, Oxtoby91, Ashcroft95} together with the symmetry
properties for the density correlation function
\eqref{twopointcrystal}. In the framework of density
functional theory the crystal is considered as an extremely inhomogeneous
distribution of matter.

To determine the inverse density correlation matrix we use the integral version of the Ornstein Zernike relation
\begin{align}
\delta (\mathbf{r_1-r_3}) &= \int d^d\!r_2 \ \langle \delta\rho(\mathbf{r_1})\delta\rho(\mathbf{r_2})\rangle \ C(\mathbf{r_2,r_3})\;,\label{OZequation}\\
C(\mathbf{r_2,r_3}) &= \frac{\delta(\mathbf{r_2-r_3})}{n(\mathbf{r_2})}- c(\mathbf{r_2,r_3})\;.\notag
\end{align}
The first term of the inverse $C(\mathbf{r_2,r_3})$ is the ideal gas contribution, whereas the second part, the direct correlation function $c(\mathbf{r_2,r_3})$, is the contribution from the excess free energy, and describes the interactions. More precisely, $c[\mathbf{r_2,r_3};n(\mathbf{r})]$ is a functional of the equilibrium density, and is obtained by the second functional derivative of the excess free energy $\mathcal{F}^{ex}$ with respect to density $n(\mathbf{r})$
\begin{equation}\label{dfndirectcorr}
c[\mathbf{r_1,r_2};n(\mathbf{r})]=\beta\frac{\delta \mathcal{F}^{ex}[n(\mathbf{r})]}{\delta n(\mathbf{r_1})\delta n(\mathbf{r_2})} = c[\mathbf{r_2,r_1};n(\mathbf{r})]\;.
\end{equation}

In the following manipulations the symmetry expressed in \eqref{twopointcrystal} is used to derive an expression for the inverse density correlation function $J_\mathbf{g g^\prime}(\mathbf{q})$ in terms of the direct correlation function $c(\mathbf{r_1,r_2})$ starting with the Ornstein Zernike (OZ) equation.
The left hand side of \eqref{OZequation} becomes
\begin{equation}
\int d^d\!r_1 \int d^d\!r_3
e^{i(\mathbf{g+q})\cdot\mathbf{r_1}}e^{-i(\mathbf{g^{\prime\prime}+q})\cdot\mathbf{r_3}}\
\delta (\mathbf{r_{13}}) = V \delta_{\mathbf{g g^{\prime\prime}}}\;.
\end{equation}
The right hand side is
\begin{align}
 & \int\!d^d\!r_1\!\int\!d^d\!r_2\!\int\!d^d\!r_3 e^{i\mathbf{(g+q)\cdot r_1}} e^{-i\mathbf{(g^{\prime\prime}+q)\cdot r_3}}\times \notag\\
 &\qquad \times \langle \delta\rho(\mathbf{r_1})\delta\rho(\mathbf{r_2})\rangle \ C(\mathbf{r_2,r_3}) \notag\\
 &=  \sum_\mathbf{G} \delta n^{(2)}_\mathbf{G}(\mathbf{-G/2 -g-q}) \ C(\mathbf{-G-g-q,g^{\prime\prime}+q})\;,
\end{align}
which, with Eq.~\eqref{symSg3}, yields
\begin{align}
V\delta_\mathbf{gg^{\prime\prime}}\!&=\!\!\sum_\mathbf{g^\prime}\langle\delta\rho^\ast(\mathbf{g+q})\delta\rho(\mathbf{g^\prime+q})\rangle
\frac{1}{V} C(\mathbf{-g^\prime-q,g^{\prime\prime}+q}).
\end{align}
So the inverse density correlation function is a special kind of Fourier transformation of essentially the direct correlation function.
\begin{align}
 &J_{\mathbf{gg^\prime}}(\mathbf{q}) = \frac{k_BT}{V} \ C(\mathbf{-g-q,g^\prime+q})\notag \\
 &=\!\!\frac{k_BT}{V}\!\!\!\int\!\!d^d\!r_1\!\!\!\int\!\!d^d\!r_2 e^{i\mathbf{g\cdot r_1}}e^{-i\mathbf{g^\prime\cdot r_2}} e^{i\mathbf{q\cdot r_{12}}}\!\!\left(\!\frac{\delta(\mathbf{r_{12}})}{n(\mathbf{r_1})}\!-\!c(\mathbf{r_1,r_2})\!\right)\!. \label{Jdfn}
\end{align}
With the definition of the direct correlation function
(\ref{dfndirectcorr}) and its symmetry under interchange of
$\mathbf{r_1}\leftrightarrow\mathbf{r_2}$ the hermitian property of
$J_\mathbf{g g^\prime}(\mathbf{q})$ follows
\begin{equation}\label{Jsymmetry}
J_\mathbf{g g^\prime}(\mathbf{q}) = J_\mathbf{-g\prime, -g}(-\mathbf{q})= (J_\mathbf{gg^\prime}(\mathbf{q}))^{\ast T}= J^\dagger_{\mathbf{gg^\prime}}(\mathbf{q})\;.
\end{equation}

\subsection{Invariance under Global Transformations}
One of the fundamental results of density functional theory \cite{Evans79,Rowlinson82,Oxtoby91,Ashcroft95} is that the external potential $V^{ext}(\mathbf{r})$ is a functional of the equilibrium density $n(\mathbf{r})$. That is, for a given equilibrium density the external potential $V^{ext}(\mathbf{r})=V^{ext}[\mathbf{r};n(\tilde{\mathbf{r}})]$ is uniquely determined.

A functional Taylor expansion thus yields
\begin{align}
\delta V^{ext}(\mathbf{r_1})\! &=\! \int d^d\!r_2 \frac{\delta V^{ext}(\mathbf{r_1})}{\delta n(\mathbf{r_2})} \delta n(\mathbf{r_2})\;.
\end{align}
As the internal state of a crystal is unaffected by a global translation and rotation, we now consider the effects of such  transformations explicitly. This derivation can also be considered an invariance principle \cite{Baus84,Lovett91}, and yields relations of the direct correlation function for a crystal. In the classical approach to elasticity, particle interaction is described via a potential \cite{Wallace70}. In that approach the consequences of invariance are the conditions for the microscopic expressions of the derivatives of the particle potential, which ensure the macroscopic Voigt symmetry of the elastic constants. We derive analogous results now using the direct correlation function.

\subsubsection{Translational Invariance}

In the case of a simple translation the transformation is given by $\mathbf{r^\prime}=\mathbf{r+s}$, and the functional Taylor expansion yields
\begin{align}
V^{ext}\!(\mathbf{r_1\!+s})\! &=\! V^{ext}\!(\mathbf{r_1}\!)\! +\!\!\!\int\!\!\!d^d\!r_2 \frac{\delta V^{ext}\!(\mathbf{r_1}\!)}{\delta n(\mathbf{r_2}\!)}\! \Big(\!n(\mathbf{r_2\!+\!s})\!-\!n(\mathbf{r_2})\!\Big).
\end{align}
For an infinitesimal translation $s_\alpha\to 0$ and with a further relation from density functional theory
\begin{equation}
\beta\frac{\delta V^{ext}(\mathbf{r_1})}{\delta n(\mathbf{r_2})} =  c(\mathbf{r_1,r_2}) - \frac{\delta(\mathbf{r_{12}})}{n(\mathbf{r_1})}\;,
\end{equation}
one obtains the Lovett, Mou, Buff \cite{Lovett76}, Wertheim \cite{Wertheim76} equation (LMBW)
\begin{equation}\label{LMBW1}
\nabla_\alpha^{(1)}\Big[\ln n(\mathbf{r_1})+\beta V^{ext}(\mathbf{r_1})\Big] = \int d^d\!r_2 c(\mathbf{r_1,r_2}) \nabla_\alpha^{(2)}n(\mathbf{r_2})\;.
\end{equation}
Without external potential the trivial solution is the one with a homogeneous equilibrium density, i.e. a fluid. As we are interested in periodic equilibrium densities, the limit of vanishing external field is taken which leads to a non-trivial solution. The right hand side is interpreted as an effective force on a particle due to interactions with the other particles \cite{Evans79}.

\subsubsection*{Constants of Elasticity}\label{sectionconstantselasticity}

In this paragraph the dependence on wave vector $\mathbf{q}$ of $\Lambda_{\alpha\beta}(\mathbf{q})$ 
in the hydrodynamic limit is derived with the help of the LMBW
equation \eqref{LMBW1}. To do so three constants of elasticity are introduced
\begin{align}
\Lambda_{\alpha\beta}(\mathbf{q}) &= \sum_{\mathbf{g^\prime,g}} (g^\prime +q )_\alpha n_\mathbf{g^\prime}^\ast J_{\mathbf{g^\prime g}}^\ast(\mathbf{q}) n_\mathbf{g}(g+q)_\beta \tag{\ref{Lambdadfn}}\\
&= \Big(\lambda_{\alpha\beta}(\mathbf{q})-iq_\alpha \mu_\beta(\mathbf{q})+iq_\beta \mu_\alpha^\ast(\mathbf{q})+q_\alpha \nu(\mathbf{q}) q_\beta\Big),
\end{align}
according to the explicit powers in $\mathbf{q}$. The matrix $\Lambda_{\alpha\beta}(\mathbf{q})$ is hermitian, which is a consequence of \eqref{Jsymmetry}.

We now discuss the three constants of elasticity separately, and start with the simplest case, which is the term proportional to
$q_\alpha q_\beta$. The realness of the equilibrium density 
$n(\mathbf{r})=\sum_\mathbf{g}n_\mathbf{g}e^{i\mathbf{g\cdot r}}=
\sum_\mathbf{g}n^\ast_\mathbf{g}e^{-i\mathbf{g\cdot r}}=
n^\ast(\mathbf{r})$ yields
\begin{subequations}
\begin{align}
\nu(\mathbf{q}) &= \sum_{\mathbf{g,g^\prime}} n_\mathbf{g^\prime}^\ast J_{\mathbf{g^\prime g}}^\ast(\mathbf{q}) n_\mathbf{g}\\
 &=\frac{k_BT}{V}\int d^d\!r_1 \int\! d^d\!r_2 n(\mathbf{r_1}) n(\mathbf{r_2}) e^{-i\mathbf{q\cdot r_{12}}}\notag\\
 &\qquad\left(\frac{\delta(\mathbf{r_{12}})}{n(\mathbf{r_1})}-c(\mathbf{r_1,r_2})\right)\\ 
 &\approx \nu + \mathcal{O}(q^2)\;,
\end{align}
where the homogeneous constant $\nu$ equals
\begin{align}
\nu &= \!\frac{k_BT}{V}\!\!\int\!\! d^d\!r_1\!\! \int\!\! d^d\!r_2 \Big(\!n(\mathbf{r_1})\delta(\mathbf{r_{12}})-n(\mathbf{r_1})c(\mathbf{r_1,r_2})n(\mathbf{r_2})\! \Big).
\end{align}
It also follows, that
\begin{align}
\nu(\mathbf{q}) &= \nu^\ast(\mathbf{q})\;,
\end{align}
\end{subequations}
is even in $q$. 
The fact that $\nu(\mathbf{q})\in \mathbbm{R}$ and that it has only
even powers in an expansion in $q$ is a consequence of the
$\mathbf{r_1}\leftrightarrow\mathbf{r_2}$ symmetry. One further
interesting fact is, that the equation for $\nu$ reduces to the
inverse compressibility $\kappa$ of a fluid\cite{Rowlinson82}
for $n(\mathbf{r})=n_0$ and $c(\mathbf{r_1,r_2})=c(r_{12})$ 
\begin{equation}
\kappa^{-1} = \frac{k_B T}{V} \Big( N - n^2_0V \int \ c(r_{12}) \ d^d\!r_{12} \Big)\;.\tag{(4.27) in [\onlinecite{Rowlinson82}]}
\end{equation}
The next term $\mu_\alpha(\mathbf{q})$ is manipulated with the help of the gradient of the equilibrium density $\nabla_\alpha n(\mathbf{r})=\sum_\mathbf{g}i g_\alpha n_\mathbf{g}e^{i\mathbf{g\cdot r}}$, and the LMBW equation in the limit of vanishing external potential
\begin{subequations}
\begin{align}
\mu_\alpha(\mathbf{q}) &= \sum_{\mathbf{g,g^\prime}} n_\mathbf{g^\prime}^\ast J_{\mathbf{g^\prime g}}^\ast(\mathbf{q}) n_\mathbf{g} i g_\alpha\\
 &= \frac{k_BT}{V} \int d^d\!r_1 \int d^d\!r_2 n(\mathbf{r_1}) \nabla_\alpha n(\mathbf{r_2}) e^{-i\mathbf{q\cdot r_{12}}}\notag\\
 &\qquad\left(\frac{\delta(\mathbf{r_{12}})}{n(\mathbf{r_1})}-c(\mathbf{r_1,r_2})\right)\\
 &\stackrel{\eqref{LMBW1}}{=} \frac{k_BT}{V} \int d^d\!r_1 \int d^d\!r_2 n(\mathbf{r_1}) \nabla_\alpha n(\mathbf{r_2}) c(\mathbf{r_1,r_2})\notag\\
&\qquad \Big( 1 - e^{-i\mathbf{q\cdot r_{12}}}\Big)\\
 &\approx i\mu_{\alpha\beta}q_\beta + \mathcal{O}(q^2)\;,
\end{align}
where the second rank tensor $\mu_{\alpha\beta}$ describing the long wavelength limit equals
\begin{align}
\mu_{\alpha\beta} &= \frac{k_BT}{V} \int d^d\!r_1 \int d^d\!r_2 n(\mathbf{r_1}) \nabla_\alpha n(\mathbf{r_2}) c(\mathbf{r_1,r_2}) r_{12,\beta}\;. 
\end{align}
\end{subequations}
For a crystal with inversion symmetry it can be shown that the correction in the expansion of $\mu_\alpha(\mathbf{q})$ is $\mathcal{O}(q^3)$. 
The realness of the gradient of the equilibrium density $\nabla_\alpha n(\mathbf{r})\in\mathbbm{R}$, i.e. $\nabla_\alpha n(\mathbf{r})=\sum_\mathbf{g}i g_\alpha n_\mathbf{g}e^{i\mathbf{g\cdot r}}=\sum_\mathbf{g} -i g_\alpha n^\ast_\mathbf{g} e^{-i\mathbf{g\cdot r}}$, together with the LMBW equation is used for the last term $\lambda_{\alpha\beta}(\mathbf{q})$
\begin{subequations}\label{3dfnlambda}
\begin{align}
\lambda_{\alpha\beta}(\mathbf{q}) &= \sum_{\mathbf{g,g^\prime}} -ig^\prime_\alpha n_\mathbf{g^\prime}^\ast J_{\mathbf{g^\prime g}}^\ast(\mathbf{q}) n_\mathbf{g} i g_\beta \\
 &= \frac{k_BT}{V}\int d^d\!r_1 \int d^d\!r_2 \nabla_\alpha n(\mathbf{r_1}) \nabla_\beta n(\mathbf{r_2}) c(\mathbf{r_1,r_2})\notag\\
&\qquad\Big( 1-e^{-i\mathbf{q\cdot r_{12}}}\Big)\\
 &\approx \lambda_{\alpha\beta\gamma\delta}q_\gamma q_\delta + \mathcal{O}(q^4)\;,
\end{align}
where the fourth rank tensor equals
\begin{align}
\lambda_{\alpha\beta\gamma\delta}\! &=\!\! \frac{k_BT}{2V}\!\!\!\int\!\!\! d^d\!r_{\!1}\!\!\! \int\!\!\! d^d\!r_{\!2}\! \nabla_{\!\alpha} n(\mathbf{r_1})\! \nabla_{\!\beta} n(\mathbf{r_2}) c(\mathbf{r_1, r_2}) r_{12,\gamma} r_{12,\delta}.
\end{align}
Obviously, one also finds
\begin{align}
\lambda_{\alpha\beta}(\mathbf{q}) &= \lambda^\ast_{\beta\alpha}(\mathbf{q})\;.
\end{align}
\end{subequations}
Again it can be shown that due to the $\mathbf{r_1}\leftrightarrow\mathbf{r_2}$ symmetry, the expansion in $q$ has only even powers, and that $\lambda_{\alpha\beta}(\mathbf{q})\in \mathbbm{R}$, so $\lambda_{\alpha\beta}(\mathbf{q})=\lambda_{\beta\alpha}(\mathbf{q})$.\\
Note also, that for the expansion to be valid, the direct correlation function $c(\mathbf{r_1,r_2})$ is assumed to be of short range in the difference vector $\mathbf{r_1-r_2}$.

To sum it up, it was shown in this paragraph that $\lim_{\mathbf{q}\to 0}\Lambda_{\alpha\beta}(\mathbf{q})$ is indeed second order in $q$, and this was derived with the LMBW equation, which is a consequence of translational invariance.
\begin{align}
\Lambda_{\alpha\beta}(\mathbf{q}) &\approx
\lambda_{\alpha\beta\gamma\delta}q_\gamma q_\delta 
+ q_\alpha \mu_{\beta\gamma} q_\gamma +q_\beta
\mu_{\alpha\gamma} q_\gamma +q_\alpha \nu q_\beta\; .
\end{align}

The second term in Sec.~\ref{subsecDFDD} with undetermined $q$ dependence was
\begin{align}
& V_\alpha({\bf q}) = \vec{v}^{\,\dagger}_\alpha\stackrel{\leftrightarrow}{J}^{\ast}\!\!\frac{\vec{n}}{n_0} =-\frac{i}{n_0} \sum_\mathbf{g^{\prime},g} (g^{\prime}+q)_\alpha n_\mathbf{g^{\prime}}^\ast J_\mathbf{g^{\prime}g}^\ast(\mathbf{q}) n_\mathbf{g} \tag{\ref{couplingc}}\\
&\qquad= \frac{1}{n_0} \Big(\mu_\alpha^\ast(\mathbf{q}) - iq_\alpha \nu(\mathbf{q})\Big)\\
&\qquad\approx  -\frac{iq_\beta}{n_0} \Big(\mu_{\alpha\beta} +\nu\delta_{\alpha\beta}\Big)\;.
\end{align}
Thus, the momentum equation \eqref{zwischen} indeed contains a term proportional to the gradient of the defect density.

\subsubsection{Rotational Invariance}
\begin{sloppypar}
As translational invariance was the reason behind the correct $q$ dependence of $\lim_{\mathbf{q\to 0}}\Lambda_{\alpha\beta}(\mathbf{q})$ and $\lim_{\mathbf{q\to0}}V_\alpha(\mathbf{q})$, the consequence of rotational invariance is now considered. It will be shown, that it yields symmetries in the indices of the constants of elasticity $\mu_{\alpha\beta}$ and $\lambda_{\alpha\beta\gamma\delta}$.\end{sloppypar}
An infinitesimal rotation is given by
\begin{equation}
\mathbf{r^\prime}=\mathbf{r}+\boldsymbol{\delta\theta}\times\mathbf{r}+\mathcal{O}(\delta\theta^2)\;.
\end{equation}
Thus the first order term of the expansion in $\delta\theta$ is
\begin{align}
\boldsymbol{\delta\theta}\!\times\!\mathbf{r_1}\!\cdot\!\nabla^{(1)} V^{ext}\!(\mathbf{r_1})\! &=\!\! \int\!\! d^d\!r_2\frac{\delta V^{ext}\!(\mathbf{r_1})}{\delta n(\mathbf{r_2})}\boldsymbol{\delta\theta}\!\times\!\mathbf{r_2}\!\cdot\!\nabla^{(2)}n(\mathbf{r_2}).
\end{align}
With invariance of the scalar triple product under cyclic permutations and an arbitrary $\boldsymbol{\delta\theta}$, one finally ends up with a rotational analog of the LMBW equation \cite{Schofield82}
\begin{align}
\mathbf{r_1}\!\times\!\nabla^{(1)} &\Big[\ln n(\mathbf{r_1})\!+\!\beta V^{ext}(\mathbf{r_1})\Big]\! \notag\\
 &=\! \int\! d^d\!r_2 c(\mathbf{r_1,r_2})\Big(\!\mathbf{r_2}\!\times\! \nabla^{(2)}n(\mathbf{r_2})\!\Big)\!.\label{LMBWrot}
\end{align}
This equation may be interpreted as a balance of effective torques in analogy to the balance of forces.

\subsubsection*{Symmetry of Constants of Elasticity}

The results for translational, Eq.~\eqref{LMBW1}, and rotational,
Eq.~\eqref{LMBWrot}, invariance can be combined to understand the
index symmetries of the constants of elasticity. The difference
$(\mathbf{r_1}\times$ $\eqref{LMBW1})-\eqref{LMBWrot}$, which is
valid for any $V^{ext}(\mathbf{r_1})$, yields
\begin{align}
\!\int\!\!\! d^d\!r_{\!2} c(\mathbf{r_1,r_2}) r_{12,\alpha}\! \nabla_{\!\beta}^{(2)}\!n(\mathbf{r_2})\! &=\!\!\! \int\!\!\! d^d\!r_{\!2} c(\mathbf{r_1,r_2}) r_{12,\beta}\! \nabla_{\!\alpha}^{(2)}\!n(\mathbf{r_2}).
\end{align}
Integrating the last equation with $\frac{k_BT}{V}\int d^d\!r_1 n(\mathbf{r_1}) $ leads to
\begin{align}
0 &= \frac{k_BT}{V} \int d^d\!r_1 n(\mathbf{r_1}) \int d^d\!r_2 c(\mathbf{r_1,r_2})\notag\\
&\quad \times\Big(\nabla_\alpha^{(2)}n(\mathbf{r_2})r_{12,\beta}- \nabla_\beta^{(2)}n(\mathbf{r_2})r_{12,\alpha}\Big)\;.
\end{align}
This is nothing but the statement that $\mu_{\alpha\beta}$ is symmetric in its indices
\begin{equation}
\Rightarrow \mu_{\alpha\beta} = \mu_{\beta\alpha}\;.
\end{equation}

In the same manner, as translational and rotational invariance led
to a symmetric matrix $\mu_{\alpha\beta}$, the  symmetry of the
indices of $\lambda_{\alpha\beta\gamma\delta}$ can be addressed. So
far it is known that
$\lambda_{\alpha\beta\gamma\delta}=\lambda_{\beta\alpha\gamma\delta}$
(consequence of symmetry of $\lambda_{\alpha\beta}(\mathbf{q})$) and
$\lambda_{\alpha\beta\gamma\delta}=\lambda_{\alpha\beta\delta\gamma}$
(symmetric combination in $\gamma\leftrightarrow\delta$ in definition \eqref{3dfnlambda}).
Repeating analogous arguments concerning the symmetry of $\mu_{\alpha\beta}$, one finds that $\lambda_{\alpha\beta\gamma\delta}$ is symmetric under the pairwise interchange $\alpha\beta\leftrightarrow\gamma\delta$ (see [\onlinecite{Walz09}] for details)
\begin{equation}
\Rightarrow \lambda_{\alpha\beta\gamma\delta} = \lambda_{\gamma\delta\alpha\beta}\;.
\end{equation}

\section{Results and Discussion}

\subsection{Summary of the Derived Equations of Motion}
Because the results for the equations of motion are spread over different sections, it appears useful to collect and list them. 

Starting with the conserved  (neglecting for simplicity energy conservation) and symmetry restoring fields, we  showed that the ansatz Eq.~\eqref{deltac} leads to an exact solution of the (for simplicity dissipation-less) \blue{Zwanzig-Mori equations} \eqref{bsheom}, if the seven ($d+1$) coarse grained fields satisfy the following (because of our use of the fluctuation dissipation theorem necessarily) linear equations; they hold for all $\bf q$ in the first Brillouin zone.

Mass density times the time-derivative of the displacement field equals the momentum density field
\begin{subequations}\label{EqsRaD}
\begin{equation}\label{ResDisUJ}
m n_0\; \partial_t\delta u_\alpha(\mathbf{q},t) = \delta j_\alpha(\mathbf{q},t) \;. \end{equation}
Density fluctuations arise because of the divergence of the displacement field and defect density fluctuations 
\begin{equation}
\delta n(\mathbf{q},t)= -i n_0 q_\alpha \delta u_\alpha(\mathbf{q},t)- \delta c(\mathbf{q},t)\;.
\end{equation}
Mass is conserved, which connects density and momentum density fluctuations
\begin{equation} \label{ConservationCoarseGrained}
m \partial_t \delta n(\mathbf{q},t) +  i q_\alpha \delta j_\alpha(\mathbf{q},t) = 0\;.
\end{equation}
Momentum density, displacement and defect density field are coupled in
\begin{align}
&\partial_t \delta j_\alpha(\mathbf{q},t)\!=\!-\Lambda_{\alpha\beta}(\mathbf{q})\; \delta u_\beta(\mathbf{q},t) - 
 V_\alpha(\mathbf{q})\; \delta c(\mathbf{q},t)\notag\\
 &=\!-\Big(\!\lambda_{\alpha\beta}(\mathbf{q})\!-\!iq_\alpha \mu_\beta(\mathbf{q})\!+\!iq_\beta \mu_\alpha^\ast(\mathbf{q})\!+\!q_\alpha \nu(\mathbf{q}) q_\beta\!\!\Big)\delta u_\beta(\mathbf{q},t)\notag\\
 &\quad- \frac{1}{n_0} \Big(\mu_\alpha^\ast(\mathbf{q}) - iq_\alpha \nu(\mathbf{q})\Big) \delta c(\mathbf{q},t) \; .\label{ResDisMomentum}
\end{align}
Also the wave equation for the momentum field holds
\begin{align}
&mn_0\partial^2_t \delta j_\alpha(\mathbf{q},t)  = - \Lambda_{\alpha\beta}(\mathbf{q}) \ \delta j_\beta(\mathbf{q},t) \notag\\
 &=\!\Big(\!\lambda_{\alpha\beta}(\mathbf{q})\!-\!iq_\alpha \mu_\beta(\mathbf{q})\!+\!iq_\beta \mu_\alpha^\ast(\mathbf{q})\!+\!q_\alpha\nu(\mathbf{q})q_\beta \Big) \delta j_\beta(\mathbf{q},t) \; .\label{ResDisWaveJ}
\end{align}
In order to recover the momentum wave equation Eq.~\eqref{ResDisWaveJ} by taking a derivative with respect time of Eq.~\eqref{ResDisMomentum} and using Eq.~\eqref{ResDisUJ}, the defect density has to be constant
\begin{equation}
\partial_t\, \delta c(\mathbf{q},t) = 0 \; .
\end{equation}
\end{subequations}
Taking the time derivative of Eq.~\eqref{ResDisUJ} and using Eq.~\eqref{ResDisMomentum}, one sees that the defect density plays the role of an inhomogeneity in the wave equation of the displacement field, which otherwise contains the identical constants of elasticity as the momentum one.

In the hydrodynamic limit, where $q\to0$, the elastic coefficients in Eqs.~\eqref{ResDisMomentum} reduce to the following expressions
\begin{subequations}
\begin{align}
\nu(\mathbf{q}) &= \nu +{\cal O}(q^2)\;,\\
\mu_\alpha(\mathbf{q}) &= i\mu_{\alpha\beta}q_\beta+{\cal O}(q^2)\;, \\
\lambda_{\alpha\beta}(\mathbf{q}) &= \lambda_{\alpha\beta\gamma\delta}q_\gamma q_\delta+{\cal O}(q^4)\;,
\end{align}
\end{subequations}
with the following symmetries
\begin{subequations}\label{SummarySymmetrymulambda}
\begin{align}
\mu_{\alpha\beta} &= \mu_{\beta\alpha}\;,\\
\lambda_{\alpha\beta\gamma\delta} &= \lambda_{\beta\alpha\gamma\delta} = \lambda_{\alpha\beta\delta\gamma} = \lambda_{\gamma\delta\alpha\beta}\;.
\end{align}
\end{subequations}
For later comparison this summary is completed by giving the momentum equation in the hydrodynamic limit
\begin{align}\label{momentumhydrolimit}
\partial_t \delta j_\alpha(\mathbf{q},t) &= 
iq_\beta\Big(\nu\delta_{\alpha\beta}+\mu_{\alpha\beta}\Big)\frac{\delta c(\mathbf{q},t)}{n_0}\notag\\
 &\quad -W_{\alpha\beta\gamma\delta}\, q_\beta q_\delta \, \delta u_\gamma(\mathbf{q},t)\;,
\end{align}
with a wave propagation matrix $W_{\alpha\beta\gamma\delta}$
\begin{align}
W_{\alpha\beta\gamma\delta} &= \lambda_{\alpha\gamma\beta\delta} + \nu \frac{1}{2}(\delta_{\alpha\beta}\delta_{\gamma\delta}+\delta_{\alpha\delta}\delta_{\beta\gamma})\notag\\
 &\quad + \frac{1}{2}\Big(\mu_{\alpha\delta}\delta_{\beta\gamma} +\mu_{\alpha\beta}\delta_{\gamma\delta} +\mu_{\gamma\delta}\delta_{\alpha\beta}  +\mu_{\gamma\beta}\delta_{\alpha\delta}  \Big)\;,
\end{align}
which is symmetric in $\alpha\leftrightarrow\gamma$ and $\alpha\gamma\leftrightarrow\beta\delta$ according to Eqs.~\eqref{SummarySymmetrymulambda}, and symmetrized in $\beta\leftrightarrow\delta$, as both indices are summed over.

\subsection{Phenomenological Theory}\label{phenomenology}
For the sake of easy comparison it appears worthwhile to summarize the results from thermodynamics and \blue{classical elasticity theory} in order to compare with our microscopic expressions. Especially of interest is to verify that our results obey the symmetry relations derived within the phenomenological approaches.
The derivation of \blue{elasticity theory} within nonequilibrium thermodynamics considering the conserved densities (mass, momentum, and energy) and the broken symmetry variable displacement field can be found in the literature \cite{Chaikin95,Fleming76,Martin72} and the result will just be quoted for the reversible, isothermal, and linearized case.

Let $u_{\alpha\beta}= \frac 12 \left( \nabla_\alpha u_\beta+ \nabla_\beta u_\alpha\right)$ be the (symmetrized) gradient tensor of the displacement field, which agrees with the strain field in the considered small deformation limit. The first law  for the free energy density (per volume) as functions of density $n$ and strain $u_{\alpha\beta}$ is
\begin{equation}
df = \mu d n + h_{\alpha\beta} d u_{\alpha\beta}\;,
\end{equation}
with chemical potential $\mu = \frac{\partial f}{\partial
n}\Big|_{u_{\alpha\beta}}$ and $h_{\alpha\beta}= \frac{\partial
f}{\partial u_{\alpha\beta}}\Big|_n$ the stress tensor at constant
density. Note that we keep the temperature $T$ constant throughout.

For the linearized equations of classical elasticity theory, one requires the isothermal free energy as an expansion around the equilibrium value $n_0$, which is given by
\begin{align}
f &= f_0 + \frac{\partial f}{\partial n}\Big|_{u_{\alpha\beta}} \delta n + \frac{\partial f}{\partial u_{\alpha\beta}}\Big|_{n} u_{\alpha\beta} + \notag\\
& \frac{1}{2}\!\Big(\frac{\partial^2 f}{\partial n^2 }\Big|_{u_{\alpha\beta}}\!\! \delta n^2\! +\!2\frac{\partial^2 f}{\partial n \partial u_{\alpha\beta}} \delta n\!\cdot\! u_{\alpha\beta}\! +\! \frac{\partial^2 f}{\partial u_{\alpha\beta}\partial u_{\gamma\delta}}\Big|_{n}\! u_{\alpha\beta}u_{\gamma\delta} \Big)\notag\\
 &+ \ldots \\
 &= f_0 + \mu^0 \delta n + h^0_{\alpha\beta}u_{\alpha\beta}\notag\\
&+ \frac{1}{2}A\left(\frac{\delta n}{n_0}\right)^2 + B_{\alpha\beta}\frac{\delta n}{n_0} u_{\alpha\beta} +\frac{1}{2} C^n_{\alpha\beta\gamma\delta}u_{\alpha\beta}u_{\gamma\delta} +\ldots
\end{align}
With the equilibrium values of the chemical potential $\mu^0$ and of the stress tensor at constant density $h^0_{\alpha\beta}$. The thermodynamic derivatives are: $A=n_0^2\frac{\partial^2 f}{\partial n^2}$  an inverse compressibility, $B_{\alpha\beta}=n_0\frac{\partial^2 f}{\partial n\partial u_{\alpha\beta}}$ a matrix of coupling constants, and $C^n_{\alpha\beta\gamma\delta}=\frac{\partial^2 f}{\partial u_{\alpha\beta}\partial u_{\gamma\delta}}$ the elastic coefficients at constant density. Due to rotational invariance, i.e. a symmetric strain field $u_{\alpha\beta}$, the thermodynamic derivatives obey certain symmetries: the matrix $B_{\alpha\beta}=B_{\beta\alpha}$ is symmetric, with up to six independent coefficients depending on crystal symmetry; and the elastic constants (additional symmetry due to definition as second derivatives) have the Voigt symmetry with a maximum of 21 independent elastic coefficients.

The equations of motion derived from microscopic starting point in the previous sections contain the defect density as  fluctuating variable. Thus it is convenient to introduce the defect density as thermodynamic variable using the connection between particle and defect density, Eq.~\eqref{deltan},
\begin{equation}
\delta n = - n_0 \; {u}_{\alpha\alpha} - \delta c\; .
\end{equation}
Changing thermodynamic variables from density to defect density gives for the free energy
\begin{align}
df &= -\mu d c + \Big(h_{\alpha\beta}-n_0 \mu \delta_{\alpha\beta}\Big) d u_{\alpha\beta}\\
 &= -\mu d c + \sigma_{\alpha\beta} d u_{\alpha\beta}\; ,
\end{align}
with $\mu(c, u_{\alpha\beta}) = -\frac{\partial f}{\partial c}\Big|_{u_{\alpha\beta}}$ and the stress tensor at constant defect density
\begin{equation}
\sigma_{\alpha\beta}(c, u_{\alpha\beta}) = \frac{\partial f}{\partial u_{\alpha\beta}}\Big|_c = \Big(h_{\alpha\beta}-n_0 \mu \delta_{\alpha\beta}\Big) \; .
\end{equation}

Based on the above thermodynamic expressions, the phenomenological
equations of motion for a crystal can be presented 
\begin{subequations}\label{1PhenoHydroEqns}
\begin{align}
\partial_t m \delta n(\mathbf{q},t) &= -i q_\alpha \delta j_\alpha(\mathbf{q},t)\;,\\
\partial_t \delta u_\alpha(\mathbf{q},t)  &= \frac{1}{mn_0} \delta j_\alpha(\mathbf{q},t)\;,\\
\partial_t \delta j_\alpha(\mathbf{q},t)   &= iq_\beta \sigma_{\alpha\beta}\;,
\end{align}
\end{subequations}
which express mass conservation, that the time derivative of the
displacement is the momentum density divided by the average mass
density, and that the (conserved) momentum density changes because
of stresses.

To obtain the desired linear equations of \blue{elasticity theory} for this
set of variables starting from Eq.~\eqref{1PhenoHydroEqns}, the
partial derivatives of the stress tensor $\sigma_{\alpha\beta}$ with
respect to the defect density $c$ and $u_{\alpha\beta}$ are
required. Straightforward differentiation and use of the expansion
of the free energy gives
\begin{align}
\sigma_{\alpha\beta} &=  \frac{\partial f}{\partial u_{\alpha\beta}}\Big|_{c} = \frac{\partial f}{\partial u_{\alpha\beta}}\Big|_{n} + \frac{\partial f}{\partial n}\Big|_{u_{\alpha\beta}}\frac{\partial n}{\partial u_{\alpha\beta}}\Big|_{c}\notag\\
&= h^0_{\alpha\beta}+B_{\alpha\beta}\frac{\delta n}{n_0}+C^n_{\alpha\beta\gamma\delta}u_{\gamma\delta}\notag\\
&\quad-n_0\delta_{\alpha\beta}\left(\mu^0+\frac{A}{n_0}\frac{\delta n}{n_0}+\frac{B_{\gamma\delta}}{n_0}u_{\gamma\delta} \right)\;.
\end{align}

Now everything is in place to state the phenomenological equations
of linearized, isothermal and dissipation-less \blue{elasticity theory}
with which to compare our microscopic results. With
the change from $\delta n$ to $\delta c$ the \blue{hydrodynamic equations
of motion} are 
\begin{subequations}
\begin{align}
\partial_t \delta c(\mathbf{q},t) &= 0\;,\\
\partial_t \delta u_\alpha(\mathbf{q},t)  &= \frac{1}{mn_0} \delta j_\alpha(\mathbf{q},t)\;,\\
\partial_t \delta j_\alpha(\mathbf{q},t)
&= i (A\delta_{\alpha\beta}-B_{\alpha\beta}) q_\beta \frac{\delta c(\mathbf{q},t)}{n_0} \notag \\
&\hspace{-13mm} -\! q_\beta q_\delta\! \Big(\!C^n_{\alpha\beta\gamma\delta}\! -B_{\alpha\beta}\delta_{\gamma\delta}\!-\delta_{\alpha\beta}B_{\gamma\delta}\!+ A\delta_{\alpha\beta}\delta_{\gamma\delta}\!\Big) \delta u_\gamma(\mathbf{q},t).\label{1PhenoHydroChaLub}
\end{align}
\end{subequations}
In the last equation the elastic constants at constant defect density appear
\begin{align}
C_{\alpha\beta\gamma\delta} &= \frac{\partial^2 f}{\partial u_{\alpha\beta}\partial u_{\gamma\delta}}\Big|_{c} \notag \\
 &= C^n_{\alpha\beta\gamma\delta} -B_{\alpha\beta}\delta_{\gamma\delta} - \delta_{\alpha\beta}B_{\gamma\delta}+ A\delta_{\alpha\beta}\delta_{\gamma\delta}\; ,
\end{align}
and a combination of thermodynamic derivatives $A\delta_{\alpha\beta}-B_{\alpha\beta}$, which are based on a Maxwell relation. A related combination also showed up in the coefficient in the elasticity equations of motion connecting the prefactors of the displacement and of the defect density field, see Eq.\eqref{momentumhydrolimit}.

Comparing the classical equations of \blue{elasticity theory} with Eq.~\eqref{momentumhydrolimit},
derived from the \blue{Zwanzig-Mori} equations in the hydrodynamic limit,
we can conclude complete agreement considering the wave vector dependence, but the issue of identifying the microscopic expressions with the elastic constants remains open.

\subsection{Identification of Elastic Constants}

So far we have shown, that $\Lambda_{\alpha\beta}(\mathbf{q})\propto
q^2$ and the symmetries of the constants of elasticity
$\lambda_{\alpha\beta\gamma\delta}$ and $\mu_{\alpha\beta}$. The
piece, which is still missing, is how these constants are related
with ''the elastic constants'' $C_{\alpha\beta\gamma\delta}$.

As a first observation, the term in front of $\delta c(\mathbf{q},t)$ in Eqs.~\eqref{momentumhydrolimit} and \eqref{1PhenoHydroChaLub} implies that the coefficient $\nu$
equals (up to a constant $C_0$) the thermodynamic derivative which was abbreviated as $A$
\begin{equation}
n_0^2\; \frac{\partial^2 f}{\partial n^2} = \nu + C_0\;.
\end{equation}
Also, the coupling of the density and strain fluctuations,
abbreviated as $B_{\alpha\beta}$, is given by the matrix $-\mu_{\alpha\beta}$
\begin{equation}
n_0\; \frac{\partial^2 f}{\partial n \partial u_{\alpha\beta}} = -\mu_{\alpha\beta} + C_0\delta_{\alpha\beta}\;.
\end{equation}

For the term in front of $\delta u_\gamma(\mathbf{q},t)$ in Eqs.~\eqref{momentumhydrolimit} and \eqref{1PhenoHydroChaLub} the indices
$\beta$ and $\delta$ are summed over. Consequently the fourth-rank tensor of wave
propagation coefficients $W_{\alpha\beta\gamma\delta}$, which was defined already symmetrized, 
has to be compared with symmetrized elastic constants
$\frac{1}{2}(C_{\alpha\beta\gamma\delta}+C_{\alpha\delta\gamma\beta})
= W_{\alpha\beta\gamma\delta}$.
This yields the relevant combination\cite{Born54,Wallace70,Kroener66} for the elastic constants $C_{\alpha\beta\gamma\delta}$ in terms of wave propagation matrices $W_{\alpha\beta\gamma\delta}$, or, respectively, constants of elasticity $\nu, \mu_{\alpha\beta}$, and $\lambda_{\alpha\beta\gamma\delta}$
\begin{align}
C_{\alpha\beta\gamma\delta} &= \Big(W_{\alpha\beta\gamma\delta}+W_{\beta\alpha\gamma\delta}-W_{\alpha\gamma\beta\delta}\Big) \notag\\
&=\! \Big(\lambda_{\alpha\gamma\beta\delta}+\lambda_{\beta\gamma\alpha\delta}-\lambda_{\alpha\beta\gamma\delta}\Big) +\!\nu\delta_{\alpha\beta}\delta_{\gamma\delta}\notag \\
&\quad+\! \mu_{\alpha\beta}\delta_{\gamma\delta}\!+\!\mu_{\gamma\delta}\delta_{\alpha\beta}\;.\label{theelasticconstants}
\end{align}
Several interesting results for this combination might be noted: the
set of three independent $\lambda_{\alpha\beta\gamma\delta}$, which
are not related via Voigt symmetry, occur in the combination for the
elastic constants. The combination of $\mu_{\alpha\beta}$ and
$\nu$ are only in pairs of the indices $\alpha\beta$ and
$\gamma\delta$; there is no Voigt symmetric term
$\mu_{\alpha\delta}\delta_{\beta\gamma}+\mu_{\beta\delta}\delta_{\alpha\gamma}+\mu_{\alpha\gamma}\delta_{\beta\delta}+\mu_{\beta\gamma}\delta_{\alpha\delta}$
or $\nu
(\delta_{\alpha\gamma}\delta_{\beta\delta}+\delta_{\alpha\delta}\delta_{\beta\gamma})$,
which, for an isotropic solid, corresponds to the combination of the
shear modulus.

The elastic constant at constant density are thus given by the matrix $\lambda$ defined in Eq.~(\ref{3dfnlambda}) via
\begin{align}
\frac{\partial^2 f}{\partial u_{\alpha\beta}u_{\gamma\delta}}\Big|_{n} &= \lambda_{\alpha\gamma\beta\delta}+\lambda_{\beta\gamma\alpha\delta}-\lambda_{\alpha\beta\gamma\delta} +C_0\delta_{\alpha\beta}\delta_{\gamma\delta}\;.
\end{align}

The derived results for the elastic constants in terms of the direct
correlation function parallel other known expressions for quantities
characterizing broken symmetries in terms of $c({\bf r}_1,{\bf
r}_2)$. The Triezenberg-Zwanzig expression\cite{Triezenberg72} for the surface tension between gas and
liquid phase of a phase separated simple system contains the
equivalent quantities as our results, namely the direct correlation
function and the average density profile. For the surface tension,
Kirkwood and Buff\cite{Kirkwood49} gave another equivalent expression in terms of the actual
interaction potential and the density pair correlation function. 
For the elastic coefficients familiar results in terms of the particle
interaction potentials can be found in the classical textbooks\cite{Wallace70,Born54}, yet
only for the case of  ideal crystals and in the limit of low
temperature where particles fluctuate little around the lattice
positions. Shortly, the connection can be established via the
symmetrized wave propagation coefficients
\begin{equation}
W_{\alpha\beta\gamma\delta} = -\frac{1}{2V}\sum_{i,j}
\Phi_{\alpha\gamma}(i,j) R_{ij,\beta}R_{ij,\delta}\;
,\label{A2limit}
\end{equation}
with $\Phi_{\alpha\gamma}(i,j) =
\nabla_\alpha^{(1)}\nabla_\gamma^{(2)} \Phi(\mathbf{R^i,R^j})$,
which contains the actual potential $\Phi$ and is evaluated at the
equilibrium positions $\mathbf{R}$. Interestingly, we can recover
Eq.~\eqref{A2limit} using the mean-spherical approximation
$c(\mathbf{r_1,r_2})=-\beta \Phi(\mathbf{r_1,r_2})$ and
appropriately coarse-graining \cite{Walz09}. However, we are not
aware of results equivalent to ours containing the actual potential
and the pair correlations functions at finite temperature in
non-ideal crystals.

\subsection{Displacement and Defect Density Fields}

After the identification of the coefficients appearing in \blue{elasticity theory}, it is worthwhile to turn to the microscopic definition of the displacement field which resulted from following the standard approach to \blue{Zwanzig-Mori equations} of broken-symmetry systems. For crystals of cubic symmetry, where  $\mathcal{N}_{\alpha\beta}=\mathcal{N}\, \delta_{\alpha\beta}$
simplifies in Eq.\eqref{firstsummation}, it is
\begin{equation}
\delta u_\alpha(\mathbf{q},t) = \frac{i}{\mathcal{N}}\sum_\mathbf{g}
n_\mathbf{g}^\ast\, g_\alpha\; \delta n_\mathbf{g}(\mathbf{q},t)\;.
\end{equation}
Importantly, this relation allows to determine the displacement field purely from measuring density fluctuations. No reference lattice is required. Thus, this formula can be applied in non-ideal crystals containing arbitrary concentrations of point defects like vacancies and interstitials. For our equilibrium considerations to apply, the point defects should be mobile and diffuse during the measurement, even though defect diffusion is neglected yet in our dissipation-less formulation. 
Equation \eqref{firstsummation} was first ingeniously formulated by Szamel and Ernst \cite{Szamel93}, who also started with the Fourier components of the density, but then changed to the usual set of hydrodynamic variables. Due to their change of hydrodynamic variables their results differ from ours in the interpretation of $\nu$ as inverse isothermal compressibility, and in the neglect of the  coupling term  $\langle \delta u^\ast_\alpha \delta n\rangle$; see the Appendix. Because of this, Szamel in a continuation paper \cite{Szamel97} concluded inconsistencies to phenomenological elasticity theory \cite{Fleming76}, which are absent in our results.

Also the result for the density fluctuation appearing in the equations of elasticity theory is noteworthy
\begin{align}
\delta n(\mathbf{q},t) &= \frac{n_0}{\mathcal{N}_0}\sum_\mathbf{g}
n_\mathbf{g}^\ast \delta
n_\mathbf{g}(\mathbf{q},t)\; .
\end{align}
This coarse-grained density fluctuation differs from the microscopic density fluctuation defined in Eq.~\eqref{DfnDeltaRho}. One of the consequences is that  the correlation function of the coarse-grained density is not simply related to the generalized structure factor $S_{\bf g}({\bf q})$ defined in terms of microscopic density fluctuations in Eq.~\eqref{2GSF}. In the limit of zero wave vector,  the structure factor at ${\bf g}=0$ reduces to the isothermal compressibility, $S_{{\bf g}=0}({\bf q})\stackrel{q\to0}{\longrightarrow} n_0^2 k_BT\kappa$ , while the coarse-grained density fluctuation function reduces to the thermodynamic derivate  $\langle \delta n^\ast(\mathbf{q}) \delta n(\mathbf{q})\rangle \stackrel{q\to0}{\longrightarrow} \frac{n_0^2 V k_BT}{\nu}= V k_BT\left(\frac{\partial^2 f}{\partial n^2}\right)^{-1}$ (for $C_0=0$). 

\section{Conclusions and Outlook}

The definition of a displacement field is central to the description  
of crystal dynamics. Yet, for non-ideal crystals containing point  
defects, it had been lacking. We provide the first systematic  
derivation of a microscopic expression for the displacement field in  
terms of density fluctuations with  wavelengths close to the  
reciprocal lattice vectors. We also find that the coarse grained density
field of elasticity theory differs from the (naively expected) small wavevector 
limit of the microscopic density. These expressions lead to microscopic  
formulae for the elastic constants of a crystal given in terms of the  
direct correlation functions. A discussion of the symmetries of the  
direct correlation functions recovers the (required) symmetries of the  
elastic coefficients for general crystals. Complete agreement with the  
phenomenological description given by the linearized, isothermal and  
dissipationless elasticity theory is achieved. 

A generalization of the  
approach to include energy fluctuations and dissipation is possible;  
as are  extensions to other broken symmetry systems, like  
quasicrystals and liquid crystals. Closure approximations for the  
direct correlation functions \cite{McCarley97} will enable quantitative  
evaluation of the derived formulae. A generalization of the theory is required for the inclusion of topological defects\cite{RajLakshmi88}, which destroy the order parameter $n_{\bf g}$. This could then be compared with the continuum theory of lattice defects \cite{Eshelby56,Kroener66}. Colloidal crystals would provide  model systems \cite{Reinke07,Pertsinidis01,Pertsinidis01a} where the theory can be tested.

\begin{acknowledgments}
We thank U.~Gasser and G.~Szamel for useful discussions. This work
was (partly) funded by the German Excellence Initiative.
\end{acknowledgments}

\appendix
\section{Conventional Set of Variables}\label{AppConSet}
For comparison we outline in this appendix how the \blue{Zwanzig-Mori equations of motion} are derived with the conventional set of
variables, i.e. the density $\delta n(\mathbf{q},t)$, the momentum
density $\delta j_\alpha(\mathbf{q},t)$, and, as broken symmetry
variable, the displacement field $\delta u_\alpha(\mathbf{q},t)$. In
order to explicitly calculate correlations containing density and
displacement field, we use their expressions in terms of the
microscopic density fields given in Eqs.~\eqref{firstsummation} and \eqref{secondsummation}.
The main result of this appendix is the interpretation of the terms
$\lambda_{\alpha\beta}(\mathbf{q})$, $\mu_\alpha(\mathbf{q})$, and
$\nu(\mathbf{q})$.

The \blue{Zwanzig-Mori equations} of motion Eq.~\eqref{1heom} with the slow
variables  $\langle \delta A_i\rangle \in \{\delta n, \delta j_\alpha,
\delta u_\alpha\}$ contain the non-vanishing matrix elements of the
Liouville operator
\begin{align}
\langle \delta j^\ast_\alpha(\mathbf{q})  \mathcal{L} \delta n(\mathbf{q})\rangle &= -q_\alpha n_0 Vk_BT\;, \label{jLncorr} \\
\langle \delta j^\ast_\alpha(\mathbf{q})  \mathcal{L} \delta
u_\beta(\mathbf{q})\rangle &= -i \delta_{\alpha\beta} V k_BT\;.
\label{jLucorr}
\end{align}
While the former macroscopically follows immediately with Eq.~\eqref{ConservationCoarseGrained} from the
equipartition theorem, Eq.~\eqref{momentumcorr}, the latter additionally requires
identifying the time derivative of the displacement as momentum
density (divided by mass density), Eq.~\eqref{ResDisUJ}. Under the assumption, that the microscopic fluctuations $\delta\rho(\mathbf{g+q})$ may be replaced by the hydrodynamic ones $\delta n_\mathbf{g}(\mathbf{q})$,
\begin{equation}
 \delta\rho (\mathbf{g+q}) \approx \delta n_\mathbf{g}(\mathbf{q})=-in_\mathbf{g}g_\alpha u_\alpha(\mathbf{q}) +
n_\mathbf{g}\frac{\delta n(\mathbf{q})}{n_0}\; ,\label{AssumptionMicroHydro}
\end{equation}
both matrix elements are rederived from Eq.~\eqref{jLrhocorr}. In the first
case this leads to
\begin{align}
\langle \delta j^\ast_\alpha (\mathbf{q})  \mathcal{L} \delta
n(\mathbf{q})\rangle &= -\frac{n_0}{\mathcal{N}_0}\sum_\mathbf{g}
n_\mathbf{g}^\ast\Big((g+q)_\alpha n_\mathbf{g} V k_BT \Big)\;,
\end{align}
and in the second case to
\begin{align}
\langle \delta j^\ast_\alpha(\mathbf{q}) \mathcal{L} \delta
u_\beta(\mathbf{q})\rangle &= -
i\mathcal{N}^{-1}_{\beta\gamma}\sum_\mathbf{g}n_\mathbf{g}^\ast
g_\gamma (g_\alpha +q_\alpha) n_\mathbf{g} Vk_BT \; .
\end{align}
Due to the definitions of $\mathcal N_{\alpha\beta}$ in Eq.~\eqref{firstsummation},
$\mathcal{N}_0$ in Eq.~\eqref{secondsummation} and the orthogonality Eq.~\eqref{SumsOrthogonal},
 both summations rederive Eqs.~\eqref{jLncorr} and \eqref{jLucorr}.

Because of time reversal symmetry the \blue{Zwanzig-Mori equations of motion} Eq.~\eqref{1heom} 
contain the following non-vanishing
isothermal correlations: the equipartition theorem for the momentum
density, Eq.~\eqref{momentumcorr}; the  correlation of the coarse-grained
density, $ \langle \delta n^\ast(\mathbf{q}) \delta n(\mathbf{q})\rangle$;
 the displacement correlation function, $\langle \delta
u^\ast_\alpha(\mathbf{q}) \delta u_\beta(\mathbf{q}) \rangle $, which is the inverse of the
dynamical matrix at constant density; and
a coupling between the displacement field and the coarse-grained density fluctuation,
 $\langle \delta u^\ast_\alpha(\mathbf{q}) \delta
n(\mathbf{q}) \rangle $.
The latter three correlations are calculated most easily together, as follows with the assumption Eq.~\eqref{AssumptionMicroHydro}.
We now consider the consequences for the two-point correlation
functions
\begin{align}
 &\langle \delta\rho^\ast(\mathbf{g+q})\delta\rho(\mathbf{g^\prime+q})\rangle = \\
 & n_\mathbf{g}^\ast n_\mathbf{g^\prime}\Big( g_\alpha g^\prime_\beta \langle u^\ast_\alpha(\mathbf{q})u_\beta (\mathbf{q})\rangle + \langle \frac{\delta n^\ast(\mathbf{q}) \delta n(\mathbf{q})}{n^2_0}\rangle\notag\\
&\quad +i g_\alpha\langle u_\alpha^\ast(\mathbf{q})\frac{\delta
n(\mathbf{q})}{n_0}\rangle -i \langle \frac{\delta
n^\ast(\mathbf{q})}{n_0}u_\beta(\mathbf{q})\rangle g^\prime_\beta
\Big)\; . \notag
\end{align}
This is then inserted in the Fourier transformation of the
Ornstein-Zernike equation \eqref{invrhocorr} where the definition of
the inverse density correlation \eqref{Jdfn} enters. With the same
manipulations as in chapter \ref{sectionconstantselasticity}, one
obtains
\begin{subequations}
\begin{align}
V\! k_BT\! &=\! -\mu_\beta(\mathbf{q}) \langle \frac{\delta n^\ast\!(\mathbf{q})}{n_0} u_\beta(\mathbf{q})\rangle + \nu(\mathbf{q}) \langle\frac{\delta n^\ast\!(\mathbf{q}) \delta n(\mathbf{q})}{n^2_0}\rangle,\\
0 \!&= \!\mu_\beta(\mathbf{q}) \langle u^\ast_\alpha\!(\mathbf{q}) u_\beta(\mathbf{q})\rangle - \nu(\mathbf{q}) \langle u^\ast_\alpha\!(\mathbf{q}) \frac{\delta n(\mathbf{q})}{n_0} \rangle,\\
0\! &= \!- \lambda_{\beta\delta}(\mathbf{q})\langle \frac{\delta n^\ast\!(\mathbf{q})}{n_0} u_\delta(\mathbf{q})\rangle\!\! +\! \mu_\beta^\ast(\mathbf{q}) \langle \frac{\delta n^\ast\!(\mathbf{q}) \delta n(\mathbf{q})}{n^2_0} \rangle\!,\\
V\! k_BT \delta_{\alpha\beta}\! &=
\!\lambda_{\beta\delta}(\mathbf{q})\langle u^\ast_\alpha\!(\mathbf{q})
u_\delta(\mathbf{q})\rangle - \mu^\ast_\beta(\mathbf{q})\langle
u^\ast_\alpha\!(\mathbf{q}) \frac{\delta n(\mathbf{q})}{n_0} \rangle.
\end{align}
\end{subequations}
From this set of equations, the three desired isothermal
correlations can be read off. The  correlation of the conserved
density or coarse-grained structure factor is given by the inverse of
 $\nu(\mathbf{q})$
\begin{subequations}
\begin{align}
 \langle \delta n^\ast(\mathbf{q}) \delta n(\mathbf{q})\rangle
&= \frac{n_0^2 V k_BT}{\nu(\mathbf{q})} \;. 
\end{align}
The displacement correlation function (at constant density) is given by the inverse of the matrix $\lambda_{\alpha\beta}$
\begin{align}
\langle\delta u^\ast_\alpha(\mathbf{q})\delta
u_\beta(\mathbf{q})\rangle = Vk_BT\left(\lambda_{\alpha\beta}(\mathbf{q})\right)^{-1}\; , 
\end{align} 
which shows that the displacement fluctuations are long-ranged, $\langle\delta u^\ast_\alpha(\mathbf{q})\delta
u_\beta(\mathbf{q})\rangle \propto q^{-2}$ for $q\to0$.
Lastly, the coupling between the displacement field and the
coarse-grained density fluctuation is given by the inverse of
$\mu_\alpha(\mathbf{q})$
\begin{align}
\langle\delta u^\ast_\alpha(\mathbf{q}) \delta n(\mathbf{q})\rangle
= \frac{-n_0 V k_BT}{ \mu_{\alpha}(\mathbf{q})}\;. 
\end{align}
\end{subequations}
Note that  Szamel \cite{Szamel97} assumes that this last correlation
is $\mathcal{O}(q)$, i.e. negligible in the hydrodynamic limit,
while we find that it grows like $1/q$ and can not be neglected.

Finally the Zwanzig-Mori equations of motion of a crystal in terms of the coarse-grained density, the momentum density, and the displacement field are
\begin{subequations}
\begin{align}
\partial_t  \delta n(\mathbf{q},t) &= -i\frac{1}{m} q_\alpha  \delta j_\alpha(\mathbf{q},t)\;,\\
\partial_t  \delta j_\alpha(\mathbf{q},t) &= -i q_\alpha \frac{\nu(\mathbf{q})}{n_0}  \delta n(\mathbf{q},t) - \lambda_{\alpha\beta}(\mathbf{q}) \delta u_\beta(\mathbf{q},t) \notag \\
 &\ + i q_\alpha \mu_\beta(\mathbf{q}) \delta u_\beta(\mathbf{q},t) \!+\! \frac{\mu^\ast_\alpha(\mathbf{q})}{n_0} \delta n(\mathbf{q},t)\;, \\
\partial_t \delta u_\alpha(\mathbf{q},t) &= \frac{1}{mn_0} \delta j_\alpha(\mathbf{q},t)\;.
\end{align}
\end{subequations}
The result is, with Eq.~\eqref{deltan}, equivalent to Eqs.~\eqref{EqsRaD}.

\end{document}